\documentclass[twocolumn]{aastex61}

\received{ }
\revised{ }
\accepted{ }

\submitjournal{ApJ}

\shorttitle{3D Simulations of Massive Stars: I. Wave Generation and Propagation}
\shortauthors{Edelmann et al.}

\usepackage{amsmath}
\usepackage{comment}
\usepackage{savesym}
\savesymbol{tablenum}
\usepackage{siunitx}
\restoresymbol{SIX}{tablenum}
\DeclareMathAlphabet{\mathitbf}{OML}{cmm}{b}{it}
\newcommand{\PD}[2]{\frac{\partial #1}{\partial #2}}
\renewcommand{\vec}{\mathitbf}
\newcommand{\bvf}{Brunt--V\"ais\"al\"a frequency}

\newcommand{\Ra}{\ensuremath{\mathrm{Ra}}}
\newcommand{\Pe}{\ensuremath{\mathrm{Pe}}}
\newcommand{\Pra}{\ensuremath{\mathrm{Pr}}}
\newcommand{\Rey}{\ensuremath{\mathrm{Re}}}
\newcommand{\msol}{\ensuremath{\mathrm{M}_\odot}}
\newcommand{\rsol}{\ensuremath{\mathrm{R}_\odot}}
\begin{document}

\title{Three-Dimensional Simulations of Massive Stars: I. Wave Generation and Propagation}

\correspondingauthor{Philipp Edelmann}
\email{philipp@slh-code.org}

\author[0000-0001-7019-9578]{P. V. F. Edelmann}
\affil{School of Mathematics, Statistics and Physics, Newcastle University, Newcastle upon Tyne, NE1 7RU, UK}

\author[0000-0002-7250-6524]{R. P. Ratnasingam}
\affil{School of Mathematics, Statistics and Physics, Newcastle University, Newcastle upon Tyne, NE1 7RU, UK}

\author[0000-0002-7950-0061]{M. G. Pedersen}
\affil{Instituut voor Sterrenkunde, KU Leuven, Celestijnenlaan 200D, 3001, Leuven, Belgium}

\author[0000-0001-7402-3852]{D. M. Bowman}
\affil{Instituut voor Sterrenkunde, KU Leuven, Celestijnenlaan 200D, 3001, Leuven, Belgium}

\author[0000-0002-5335-4991]{V. Prat}
\affil{AIM, CEA, CNRS, Université Paris-Saclay, Universit\'e Paris Diderot, Sorbonne Paris Cité, F-91191 Gif-sur-Yvette, France}

\author[0000-0002-2306-1362]{T. M. Rogers}
\affil{School of Mathematics, Statistics and Physics, Newcastle University, Newcastle upon Tyne, NE1 7RU, UK}
\affil{Planetary Science Institute, 1700 East Fort Lowell, Suite 106, Tucson, Arizona 85721, USA}

\begin{abstract}
  We present the first three-dimensional (3D), hydrodynamic simulations of the core convection zone (CZ) and extended radiative zone spanning from 1\% to 90\% of the stellar radius of an intermediate mass ($3\,\msol$) star. This allows us to self-consistently follow the generation of internal gravity waves (IGWs) at the convective boundary and their propagation to the surface. We find that convection in the core is dominated by plumes. The frequency spectrum in the CZ and that of IGW generation is a double power law as seen in previous two-dimensional (2D) simulations.
  The spectrum is significantly flatter than theoretical predictions using excitation through Reynolds stresses induced by convective eddies alone.
  It is compatible with excitation through plume penetration.
  An empirically determined distribution of plume frequencies generally matches the one necessary to explain a large part of the observed spectrum.
  We observe waves propagating in the radiation zone and excited standing modes, which can be identified as gravity and fundamental modes. They show similar frequencies and node patterns to those predicted by the stellar oscillation code GYRE\@. The continuous part of the spectrum fulfills the IGW dispersion relation. A spectrum of tangential velocity and temperature fluctuations close to the surface is extracted, which are directly related to observable brightness variations in stars. Unlike 2D simulations we do not see the high frequencies associated with wave breaking, likely because these 3D simulations are more heavily damped.
\end{abstract}

\keywords{hydrodynamics - stars: interiors - convection - waves}

\section{Introduction} \label{sec:intro}
In addition to sound waves, fluid dynamical systems can have other wave-like solutions for which the restoring force is not pressure but buoyancy. These waves are commonly referred to as internal gravity waves (IGWs) to distinguish them from surface gravity waves. They occur in many stratified systems, such as atmospheres and oceans, in many of which they have an important impact on the large scale dynamics. IGWs excited by equatorial convection were found to be crucial in driving the quasi-biennial oscillation (QBO) in the Earth's equatorial stratosphere \citep{baldwin2001a}. In the oceans, IGWs excited through the surface wind or tides cause turbulent mixing when they break \citep{munk1998a}.

In stars IGWs have been suggested to play an important role in angular momentum transport and chemical mixing in radiative regions, where other mechanisms are not efficient. \citet{press1981a} suggested that IGWs in the sun can cause mixing in the convectively stable interior and affect the effective radiative opacity by a factor of two or more. IGW mixing was also suggested as the cause of lithium depletion in F~stars \citep{garcia-lopez1991a} and in the sun \citep{schatzman1993a,montalban1994a,talon2005a}.

IGWs are candidates for being the cause of some of the observed properties of stars that are poorly explained by current stellar models, such as the internal rotation structure of stars \citep{beck2012a,aerts2017b}, stellar cores counter-rotating to their envelopes \citep{triana2015a,rogers2015a}, or the enhanced mass loss needed to explain certain classes of core-collapse supernovae \citep{quataert2012a}. Photometric observations suggest the presence of convectively generated IGWs in, at least, some massive stars since the observed velocity spectrum at the surface compares well to that obtained using numerical simulations of IGWs \citep{aerts2015a,bowman2019a}.

To understand the role IGWs play in all these physical situations it is important to know what spectrum of waves in frequency and wave number space is excited by convection. Theoretical work characterizing these spectra mostly focuses on two mechanisms, excitation through the Reynolds stresses of convective eddies or through penetration of plumes. The former approach was taken by \citet{lighthill1952a}, \citet{goldreich1990a}, \citet{kumar1999a}, and later by \citet{lecoanet2013a}. All these studies found a power law dependence in frequency, i.e.\ proportional to $f^{-\alpha}$, with wave frequency~$f$ and exponent~$\alpha$. The exact value of the exponent depended on the profile of the \bvf{} at the convective boundary (CB). The spectrum generated by plume penetration was first studied by \citet{townsend1966a} in a terrestrial context and later extended to stars by \citet{montalban2000a}. A recent semianalytical model for the IGW flux caused by plumes at the base of a convection zone, as is the case in the sun, has been developed by \citet{pincon2016a}. The predicted spectrum takes a very similar functional form in all these plume-driven cases, which is proportional to $\exp[-(f/f_\text{b})^2]$, with wave frequency~$f$ and the plume frequency~$f_\text{b}$.

Multidimensional hydrodynamic simulations generally do not impose a specific IGW generation mechanism, and are able to follow convection, IGW generation and propagation directly from the basic equations. Yet numerical limitations and the extreme scales within stellar interiors often restrict them to a more dissipative regime than is realistic in stars. Nevertheless, careful choice of parameters and interpretation of the results allow us to assess theoretical predictions. Simulations showed that the Li depletion in the sun cannot be explained by IGWs \citep{rogers2006a}. Similarly the uniform rotation of the sun's radiative interior is not completely caused by IGWs \citep{rogers2005b,denissenkov2008a}. \citet{rogers2013a} performed two-dimensional (2D) hydrodynamic simulations of IGW generation at the boundary of convective cores of massive stars. They found that the IGW generation spectrum is generally much shallower than theoretical predictions. It shows two frequency regimes with different slopes, suggesting different excitation mechanisms at work.
Recent research on breaking of IGWs in the radiative envelopes of massive stars affirmed the importance of the shape of the wave generation spectrum \citep{ratnasingam2019a}.

\citet{browning2004a} performed simulations of the inner 30\% in radius of a $2\,\msol$ star with methods very similar to the ones used in this work. Their work focused on convective motions in the core, overshooting, and the influence of rotation. They do not study IGWs in detail, but mention their excitation at the convective boundary. Later work by the same group included magnetic fields and specifically studied the dynamo in the convective core \citep{brun2005a}. In contrast, our work specifically studies IGW excitation and propagation and therefore includes a much larger part of the radiation zone (up to 90\% in radius).

In their work on IGWs in solar-like stars, \citet{alvan2014a} performed a detailed analysis of wave excitation and propagation, similar to the one carried out in our work. The main difference is their work is based on solar-like stars with a convective envelope and radiative core, while the opposite is the case in our $3\,\msol$ star. Propagation through a radiative envelope along a falling density gradient causes wave amplification, which makes nonlinear behavior more likely in intermediate-mass and massive stars.

The remaining parts of this paper are structured as follows: Section~\ref{sec:method} describes the hydrodynamic equations solved and their pseudo-spectral discretization. Section~\ref{sec:simulations} discusses the stellar models used as the background state of the simulations and assumptions on heating and dissipation needed for numerical reasons. The general properties of three-dimensional (3D) convection in the core are presented in Sect.~\ref{sec:convection-zone}. Frequency spectra of core convection and their implications for the generation of IGWs are discussed in Sect.~\ref{sec:conv-spectra}. The properties of the overshooting region is the subject of Sect.~\ref{sec:overshoot}. Section~\ref{sec:radiation-zone} treats IGW propagation and the modes excited in the radiation zone, while Sect.~\ref{sec:igw-surface} discusses the signature they are expected to leave on the surface, which could be observed by photometry and spectroscopy. We conclude in Sect.~\ref{sec:conclusions}.

\section{Computational Method}
\label{sec:method}
The simulations presented here are a logical continuation of those of \citet{rogers2013a}. One caveat of their work is the restriction to 2D geometry, which is expected to yield significantly different behavior of turbulence and also altered wave propagation to some degree due to the difference between 2D annulus geometry and a 3D sphere. We extend their method to 3D by using the same type of anelastic approximation, but discretizing the horizontal part of the equation in terms of spherical harmonics instead of $\sin$ and $\cos$ functions.

We solve the following equations for the deviation from the reference state (indicated by a bar, e.g., $\overline{\rho}$) given by a hydrostatic stellar evolution model,
\begin{align}
\label{eq:anelastic-rho}
\nabla \cdot \overline{\rho} \vec{v} = 0,
\end{align}
\begin{align}
\label{eq:anelastic-v}
\PD{\vec{v}}{t}&= - (\vec{v} \cdot \nabla) \vec{v}
- \nabla P - C \overline{g} \vec{\hat{r}} + 2(\vec{v} \times \vec{\hat{z}} \Omega) \\
\nonumber
& + \overline{\nu} \left( \nabla^2 \vec{v} + \frac{1}{3} \nabla (\nabla \cdot \vec{v}) \right),\\
\label{eq:anelastic-T}
\PD{T}{t}&=  - (\vec{v} \cdot \nabla) T + (\gamma - 1) T h_\rho v_r\\
\nonumber
&- v_r \left( \PD{\overline{T}}{r} - (\gamma - 1) \overline{T} h_\rho \right) + \frac{\overline{Q}}{c_v \overline{\rho}}\\
\nonumber
& + \frac{1}{c_v\overline{\rho}} \nabla \cdot (c_p \overline{\kappa}\overline{\rho}\nabla T) + \frac{1}{c_v\overline{\rho}} \nabla \cdot (c_p \overline{\kappa}\overline{\rho}\nabla \overline{T}).
\end{align}
Here, $\vec{v}$ is the 3D fluid velocity, $v_r$ its radial component, $\overline{\rho}$ is the background density, $\gamma$ is the adiabatic index of the ideal gas equation of state, $\overline{T}$ and $T$ are the temperature background and fluctuation, $\overline{\kappa}$ and $\overline{\nu}$ are the thermal and viscous diffusivities, $\overline{Q}$ is the energy release rate, $c_v$ is the specific heat at constant density, $\overline{g}$ is gravitational acceleration, and $h_\rho=\partial \ln \overline{\rho} / \partial r$ is the negative inverse of the density scale height.
We use a standard spherical coordinate system with radius~$r$, colatitude~$\theta$, and azimuthal angle~$\phi$. The unit vector $\vec{\hat{r}}$ points in radial direction. Rotating stars are set up using a rotating frame of reference with an angular velocity~$\Omega$ and rotation axis along $\vec{\hat{z}}$ in direction of the pole at $\theta=0$.

This formulation of the anelastic equations includes self-gravity perturbations~$\Phi$ to the reference state gravitational potential $\overline{\Phi}$ by introducing the reduced pressure~$P=p/\overline{\rho}+\Phi$ and co-density~$C$ \citep{braginsky1995a,rogers2005b}. This introduces no additional computational effort as long as the thermodynamic pressure is not calculated. The co-density takes the form
\begin{equation}
  \label{eq:codensity}
  C=-\frac{1}{\overline{T}}\left(T  + \frac{1}{\overline{g} \overline{\rho}}\PD{\overline{T}}{z}p \right).
\end{equation}

In their comparison of different variants of the anelastic approximation \citet{brown2012a} also investigated this variant of the anelastic equations\footnote{They call this set of equations the RG equations.}. They find that the equations do not conserve energy for non-isothermal stratifications and suggest the removal of the term proportional to $p$ in Eq.~\eqref{eq:codensity} to ensure energy conservation. While all simulations used for the analysis in Sect.~\ref{sec:results} did not contain the \citet{brown2012a} modification, we performed several test calculations including it. The modes we find in the radiation zone are not affected by the inclusion or exclusion of this factor.

\citet{brown2012a} also predict that IGW frequencies are larger by a factor of $\sqrt{\gamma}$. When comparing the frequencies generated in our simulations to those generated with the 1D pulsation code GYRE \citep{townsend2013a}, we find small deviations, initially of the order of a few \si{\micro Hz} ($\sim$~2\% relative deviation) getting larger at higher wavenumbers, but this $\sqrt{\gamma}$~factor does not explain the differences. Hence, we are unsure how this factor manifests itself in our simulations or how that work extends to these fully nonlinear simulations.

The numerical solution method we choose is similar to the approach taken by \citet{glatzmaier1984a} and in the ASH code \citep{clune1999a} with some different choices adapted to the application at hand. To implicitly fulfill Eq.~\eqref{eq:anelastic-rho} we replace the mass flux~$\overline{\rho} \vec{v}$ by its decomposition into a poloidal ($W$) and toroidal stream function ($Z$). These are related to the mass flux by
\begin{equation}
  \label{eq:pol-tor-decomp}
  \overline{\rho} \vec{v} = \nabla \times \nabla \times W \vec{\hat{r}} + \nabla \times Z \vec{\hat{r}}.
\end{equation}
Resulting purely from a curl of a vector this is naturally divergence free. Together with temperature~$T$ and reduced pressure~$P$ these form the four unknown quantities we are solving for. They are expressed as a linear combination of spherical harmonics $Y_{l,m}$ and radius-dependent, complex coefficients. For temperature this is
\begin{equation}
T(r, \theta, \phi, t) = \sum_{m=-m_\mathrm{max}}^{m_\mathrm{max}} \sum_{l=|m|}^{l_\mathrm{max}} T_{l,m}(r,t) Y_{l,m}(\theta,\phi),
\end{equation}
and its equivalent for the other quantities. This allows us to compute horizontal derivatives via computationally inexpensive recursion relations and it avoids the singularities at the poles. As the coefficients of real-valued quantities fulfill
\begin{equation}
T_{l,-m} = (-1)^m T_{l,m}^*,
\end{equation}
only the components with $m\geq0$ need to be stored. The choice of \emph{triangular truncation} ($l_\mathrm{max} = m_\mathrm{max}$) results in uniform angular resolution. The method we use is pseudo-spectral, i.e.\ the linear terms are computed in spectral space and the nonlinear terms are computed in grid space. This approach makes it necessary to chose the number of grid points in latitudinal and longitudinal direction, $N_\theta$ and $N_\phi$, corresponding to the number of spectral modes. To avoid aliasing errors we set $N_\theta = (3l_\mathrm{max} + 1)/2$ and $N_\phi = 2 N_\theta$. Details on this kind of spectral discretization can be found in \citet{glatzmaier2013a}. We do not use a spectral basis in the radial direction to be able to easily adjust the grid to the underlying stellar model. In the present case we have an increased radial resolution in the convection zone. Radial derivatives are computed using second-order finite differences accounting for the nonuniform grid spacing.

We use the implicit Crank--Nicolson method for the linear diffusion terms to avoid the strict CFL condition that depends quadratically on the step size associated with explicit time-stepping. The nonlinear terms are calculated using the explicit Adams--Bashforth linear multistep method, which makes the method second-order accurate in time. We choose a constant time step of \SI{1}{s}, which is well below the CFL condition of the explicit terms and makes later Fourier analysis of the time series easier.

\begin{figure}
  \includegraphics{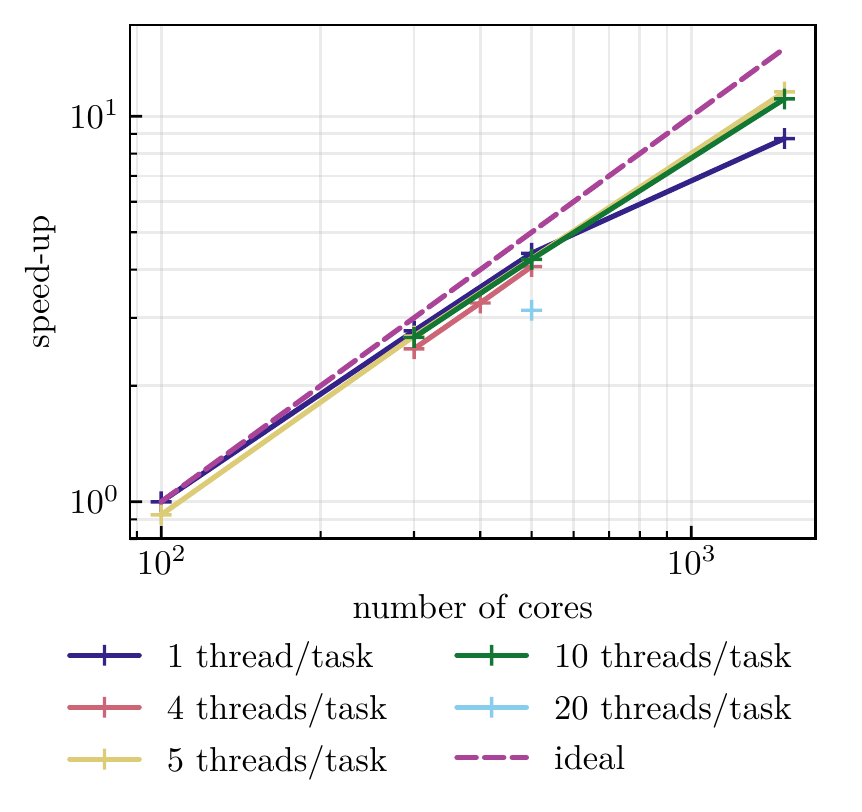}
  \caption{\label{fig:scaling-openmp}Strong scaling on the NASA NAS Pleiades system using Ivy Bridge CPUs. The reference for measuring speed-up is the case of 100~MPI tasks with 1 thread/task. The best efficiency at 1500~cores is 77\% using 5 OpenMP threads per MPI~task.}
\end{figure}

The code is parallelized using a domain decomposition in the radial coordinate only. The communication involves halo updates for computing finite differences in the radial direction and all-to-all communication for solving the linear equations involved in implicit time-stepping. It is implemented using the message passing interface (MPI). To alleviate the problem that domains become small when using many cores we additionally implement thread-based parallelization using OpenMP, which starts to be more efficient than pure MPI when there are less than 3 radial points per task (see Fig.~\ref{fig:scaling-openmp}). The achieved scaling efficiency from 100 to 1500~cores is 77\% on the NASA Pleiades system.

\section{Simulations} \label{sec:simulations}
The equations discussed in Sect.~\ref{sec:method} rely on a spherically symmetric reference state for the thermodynamic variables on top of which the evolution of small perturbations is calculated. We use the MESA (Modules for Experiments in Stellar Astrophysics) stellar evolution code\footnote{The MESA version used was SVN revision number 10000.} \citep{paxton2011a,paxton2013a,paxton2015a,paxton2018a} to produce the reference state. We use the default settings to generate a nonrotating, $3\,\msol$ zero-age main-sequence (ZAMS) star of metallicity $Z=10^{-2}$. The exact code configuration (inlists) and MESA profiles can be obtained at this URL\footnote{\url{https://www.mas.ncl.ac.uk/~npe27/igw3d/}}. No convective overshooting was used. The values of density, temperature, and gravity are adopted unchanged from the model and interpolated onto a grid with 400~cells in the convection zone and 1100~cells in the radiation zone. The total radius of the star is $R_\star = \SI{1.42e11}{cm} = 2.05\,\rsol$.

\begin{figure}
  \includegraphics{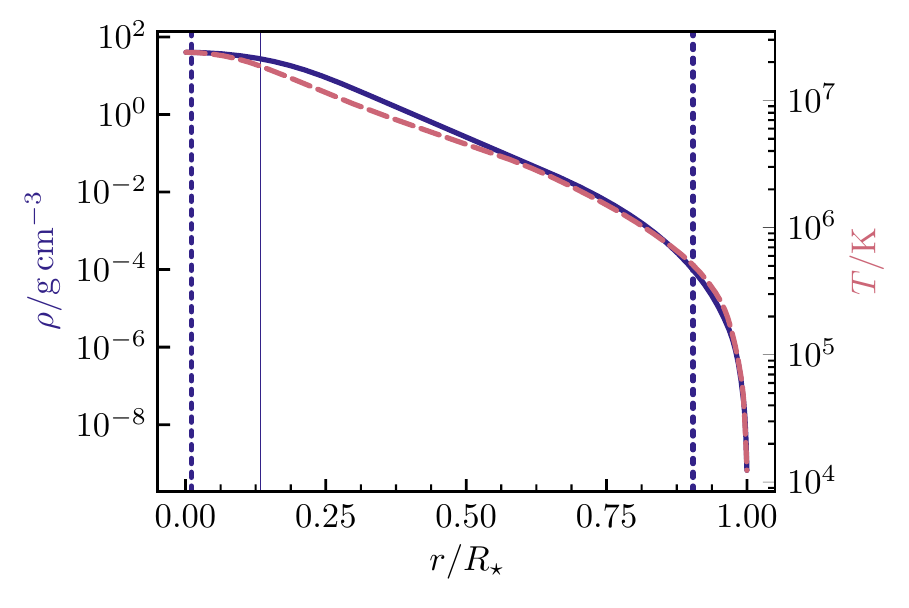}
  \caption{\label{fig:rho-t-mesa}Background stratification of density~$\rho$ (\emph{solid blue line}) and temperature~$T$ (\emph{dashed red line}) used in the 3D anelastic simulations. The vertical dotted lines show the extent of the simulation domain. The radius coordinate is scaled to the total radius of the star $R$. The vertical solid line indicates the convective--radiative boundary.}
\end{figure}

\begin{figure}
  \includegraphics{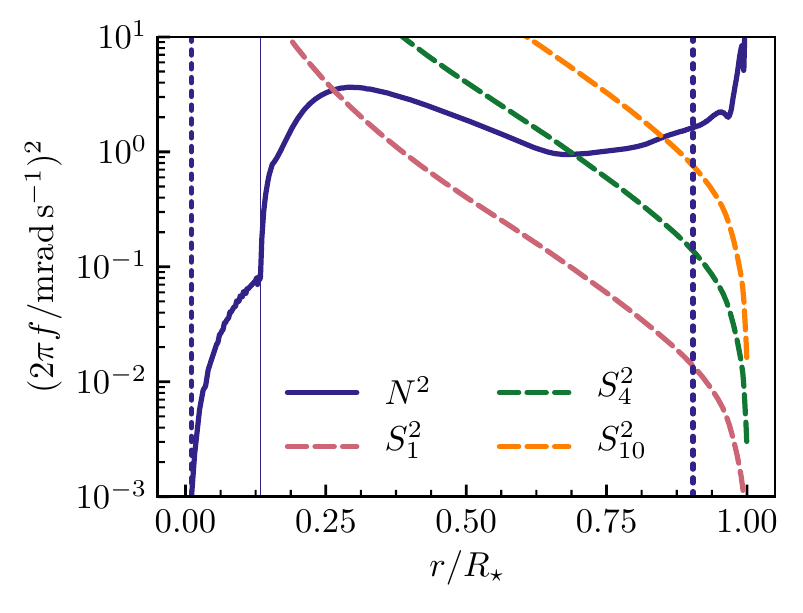}
  \caption{\label{fig:brunt-mesa}Square of \bvf{}~$N^2$ and Lamb frequencies~$S^2_l$ (dashed) for the background stratification from Fig.~\ref{fig:rho-t-mesa}. The vertical solid line indicates the outer boundary of the core convection zone and the vertical dashed lines are the boundaries of the computational domain of the 3D simulations.}
\end{figure}

Figure~\ref{fig:rho-t-mesa} shows the density and temperature profile of the stellar model. The radial extent of the 3D~simulation domain indicated by vertical, dashed lines is limited at 1\% of the stellar radius to avoid the coordinate singularity at the core and at 90\%, where density drops below $10^{-4}\,\mathrm{g\,cm^{-3}}$, covering nearly six orders of magnitude in density.\footnote{Models~H7E and H7E-HR were run with an earlier version of the stellar model, which reached 95\% of stellar radius. There is no qualitative change in the wave spectra of these models.} The \bvf{} profile, which governs the propagation of IGWs, is plotted in Fig.~\ref{fig:brunt-mesa}. It shows the convective radiative boundary at 13\% of the total radius.

\begin{deluxetable*}{c|ccccccccc}
  \tablecaption{\label{tab:models}List of 3D simulations}
  \tablehead{\colhead{Model\tablenotemark{a}} & \colhead{spectral modes} & \colhead{$\nu/\SI{e13}{cm^2.s^{-1}}$} & \colhead{$\kappa$} & \colhead{$\overline{Q}$} & \colhead{$\Omega/\SI{e-6}{rad.s^{-1}}$} & \colhead{Ra\tablenotemark{f}} & \colhead{Re\tablenotemark{f}} &\colhead{Pr\tablenotemark{f}} & \colhead{$t_\mathrm{sim}$\tablenotemark{g}/d}}
\startdata
H6R5   & 3741\tablenotemark{b}  & $8$ (rising\tablenotemark{d})    & $10^5 \kappa_\star$                                       & $10^6 \varepsilon_\star$                & $5$  & \num{8e11}                 & 126 & 60 -- \num{0.02}\tablenotemark{e} & $38.7$ \\
H6R10  & 3741\tablenotemark{b}  & $8$ (rising\tablenotemark{d})    & $10^5 \kappa_\star$                                       & $10^6 \varepsilon_\star$                & $10$ & \num{8e11}                 & 126 & 60 -- \num{0.02}\tablenotemark{e} & $59.3$ \\
H5     & 3741\tablenotemark{b}  & $1$     & $10^5 \kappa_\star$                                       & $10^5 \varepsilon_\star$                & 0    & \num{6e11}                & 468 & 7 -- \num{e-6}\tablenotemark{e}   & $13.3$ \\
H6E    & 3741\tablenotemark{b}  & $10$    & $10^5 \kappa_\star$ -- $50 \kappa_\star$\tablenotemark{e} & $\approx 10^6 \varepsilon_\star$ (exp.) & 0    & \num{e12}                 & 101 & 100 -- 2\tablenotemark{e}         & $61.7$ \\
H6LD   & 3741\tablenotemark{b}  & $10$    & $10^5 \kappa_\star$ -- $50 \kappa_\star$\tablenotemark{e} & $10^6 \varepsilon_\star$                & 0    & \num{e12}                 & 100 & 100 -- 2\tablenotemark{e}         & $58.8$ \\
H6LD-HR& 14706\tablenotemark{c}  & $10$    & $10^5 \kappa_\star$ -- $50 \kappa_\star$\tablenotemark{e} & $10^6 \varepsilon_\star$                & 0    & \num{e12}                 & 100 & 100 -- 2\tablenotemark{e}         & $6.5$ \\
H7E    & 3741\tablenotemark{b}  & $8$ (rising\tablenotemark{d})    & $\SI{5e13}{cm^2.s^{-1}}$                                  & $\approx 10^7 \varepsilon_\star$ (exp.) & 0    & \num{2e9}                 & 272 & 1 -- 39\tablenotemark{e}          & $16.4$ \\
H7E-HR & 14706\tablenotemark{c} & $8$ (rising\tablenotemark{d})    & $\SI{5e13}{cm^2.s^{-1}}$                                  & $\approx 10^7 \varepsilon_\star$ (exp.) & 0    & \num{2e9}                 & 272 & 1 -- 39\tablenotemark{e}          & $33.2$ \\
\enddata
\tablenotetext{a}{Model names are built up from the luminosity boosting factor, rotation rate, and angular resolution.}
\tablenotetext{b}{This corresponds to angular resolution of 128 ($\theta$) and 256 ($\phi$).}
\tablenotetext{c}{This corresponds to angular resolution of 256 ($\theta$) and 512 ($\phi$).}
\tablenotetext{d}{The value given is the one in the core. In the envelope $\nu$ rises with $\nu\propto\rho^{-1/4}$. The profile is continuous.}
\tablenotetext{e}{The first value applies to the core, the second to the envelope.}
\tablenotetext{f}{The definitions are given in Eqs.~\eqref{eq:rayleigh} -- \eqref{eq:prandtl}.}
\tablenotetext{g}{This is the total physical runtime of the simulation.}
\end{deluxetable*}

For numerical stability we need to increase the thermal diffusivity~$\kappa$ and kinematic viscosity~$\nu$ beyond their physical values in the star, $\kappa_\star$ and $\nu_\star$, respectively. In the $3\,\msol$ MESA model $\kappa_\star$ ranges from \SI{e7}{cm^2.s^{-1}} in the CZ to \SI{e12}{cm^2.s^{-1}} at the top of the simulated region ($r=0.9\,R_\star$), and $\nu_\star$ ranges from \SI{60}{cm^2.s^{-1}} to \SI{5e4}{cm^2.s^{-1}}. As increased diffusivity and viscosity would damp convection too strongly in order to reach a somewhat turbulent state, we increase the luminosity of the star by a similar factor to balance the increased damping. In a series of models we explore the effect of increased forcing and that of using different profiles for viscosity. These are summarized in Table~\ref{tab:models}. To put this into context we compare several characteristic nondimensional numbers. The Rayleigh number,
\begin{equation}
  \label{eq:rayleigh}
  \Ra  = \frac{g \overline{Q} D^5}{c_v \kappa^2 \nu \overline{T}},
\end{equation}
with a typical length scale~$D$ (chosen to be the size of the convective core in this case), controls the details of convection and determines if energy transport is mostly through radiation or convection. This particular form of $\Ra$ is also called a flux Rayleigh number. The stellar value is \num{e28}, which is more than six orders of magnitude higher than the values reached in the simulations. This is the rationale for increasing the energy release. If we had used the original value of $\overline{Q}$, \Ra{} would be approximately \num{e6}, which might even be subcritical. The actual convection in the star is likely even more vigorous and plume dominated than that observed in the simulations.

Flows with a high Reynolds number,
\begin{equation}
  \label{eq:reynolds}
  \Rey = \frac{v_\text{rms} D}{\nu},
\end{equation}
develop turbulence, while low values of \Rey{} normally result in laminar flow. Due to the extreme length scales~$D\approx 14\%\,R_\star$ and velocities, \Rey{} is typically extremely large in stellar environments, in the current case $\Rey\approx\num{e12}$. These parameters are not currently possible in numerical simulations. As can be seen in Table~\ref{tab:models} we can only reach values of approximately \num{e2} in the CZ\@. However, it is expected that as long as a part of the inertial range of the turbulent cascade is numerically resolved, the energy dissipation rate will not change significantly at higher \Rey{} \citep[e.g.,][Chapter~5]{frisch1995a}. Yet the small scale velocity field will definitely show differences, which is a caveat of the presented simulations. The Reynolds number is another reason for using an increased convective forcing, as using the original value would result in velocities corresponding to $\Rey \approx 1$.

The Prandtl number,
\begin{equation}
  \label{eq:prandtl}
  \Pra = \frac{\nu}{\kappa},
\end{equation}
is the ratio of viscous to thermal diffusion. In stars it is typically extremely low, ranging from \num{e-6} in the core to \num{e-9} at the surface. The only way to reach these values in our numerical simulations would be to increase $\kappa$ to very high values, which would damp the waves too much. As a compromise we settle on \Pra{} around 1 in the envelope and around 100 in the core in most models (see Tab.~\ref{tab:models}). In a few models \Pra{} reaches much lower values of \num{0.02} or even \num{e-6} in the envelope, but these are subject to excessive damping due to too much thermal diffusion.

In most models we increase luminosity by setting the heating function $\overline{Q}$ to the nuclear energy generation rate from MESA multiplied by a constant factor. These models are referred to with a name starting with ``H$X$'' for $10^X$ times the stellar luminosity~$L_\star$. For example, ``H6'' corresponds to a luminosity of $10^6\,L_\star$. In a few models we used an exponential heating function,
\begin{equation}
  \overline{Q} = A c_v \overline{\rho} \exp(-r/r_\mathrm{min}) (r-r_\mathrm{min})/R_\star,
\end{equation}
with a scaling factor $A$, which is used to adjust it to a boosted stellar luminosity. These are labeled with ``H$X$E'' for an exponential heating profile corresponding to a luminosity of $10^X$ the stellar value, and $r_\mathrm{min}$ the innermost radius of the simulation domain.

The thermal diffusivity is treated in a similar way by multiplying the stellar value with a constant factor, which was the lowest value that did not show stability problems. As the increased diffusivity is mainly needed in the convection zone, we also tried a different approach where just the CZ is subject to a value of $10^5 \kappa_\star$, while diffusivity in the radiation zone can be reduced to $50\kappa_\star$. Both regions are blended using a hyperbolic tangent function with a width of \SI{5e9}{cm} (3.5\% of the stellar radius and 26\% of the size of the convection zone). This was used in models~H6E and H6LD\@. Model~H6LD is a combination of the low diffusivity of Model~H6E with the boosted MESA energy release of model H6\@.

Important conclusions in this paper are drawn from spectra. To clarify their interpretation we give an exact definition here. For a real function~$E(t)$ sampled in an interval $[t_a, t_b]$ the Fourier transform is
\begin{equation}
  \label{eq:FT}
  \hat{E}(f) = \frac{1}{t_b - t_a} \int_{t_a}^{t_b} E(t) e^{- 2 \pi i f t} dt.
\end{equation}
By normalizing with the length of the interval the units of $\hat{E}$ are the same as those of $E$, which makes it easier to interpret the magnitude of components of the spectra.

As data from the simulations is sampled at discrete times $t_0,\ldots,t_{n-1}$ with equidistant spacing~$\Delta t$, we approximate Eq.~\eqref{eq:FT} with a discrete Fourier transform (DFT),
\begin{equation}
  \label{eq:DFT}
  \hat{E}(f_j) = \frac{1}{n} \sum_{m=0}^{n-1} E(t_m) e^{-2 \pi i \frac{{m j}}{n}},
\end{equation}
where $j$ takes values from 0 to $\lfloor\frac{n}{2}\rfloor$. Higher values of $j$ are redundant due the real input data. The corresponding frequencies are $f_j = \frac{j}{n\Delta t}$.

\section{Results} \label{sec:results}
\subsection{Convection Zone}
\label{sec:convection-zone}
\begin{figure}
  \includegraphics{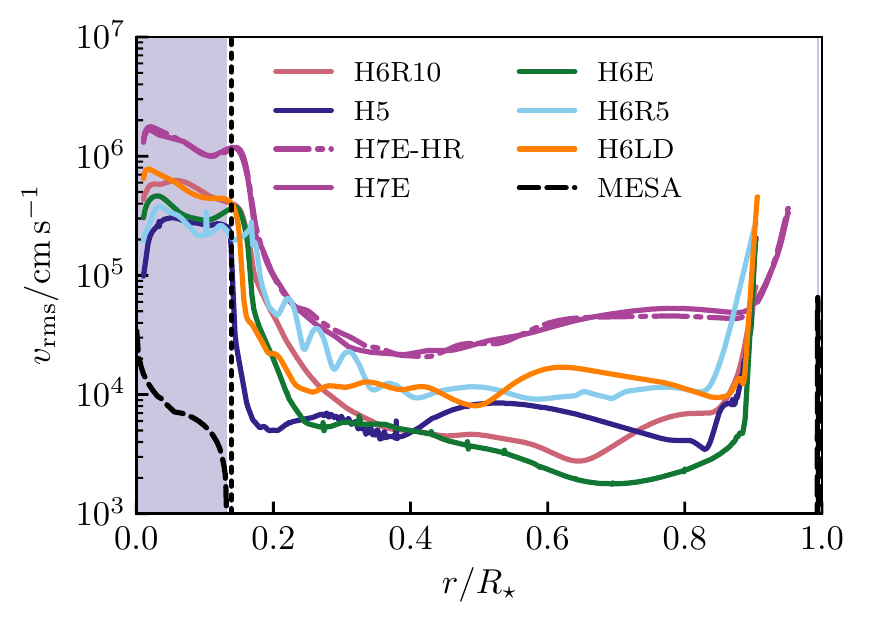}
  \caption{\label{fig:convvel-compare}Root mean square velocity as a function of radius in the 3D simulations for different luminosity boosting factors. The dashed line is the velocity estimate according to mixing-length theory returned from MESA\@. The surface convection zone is visible at $r\approx R_\star$ in the MESA data. The region shaded in blue marks the position of the convection zones in MESA\@. The vertical dashed line is the radius at which the spectra from Fig.~\ref{fig:CZ-spec-all} are computed.}
\end{figure}
\begin{figure}
  \includegraphics{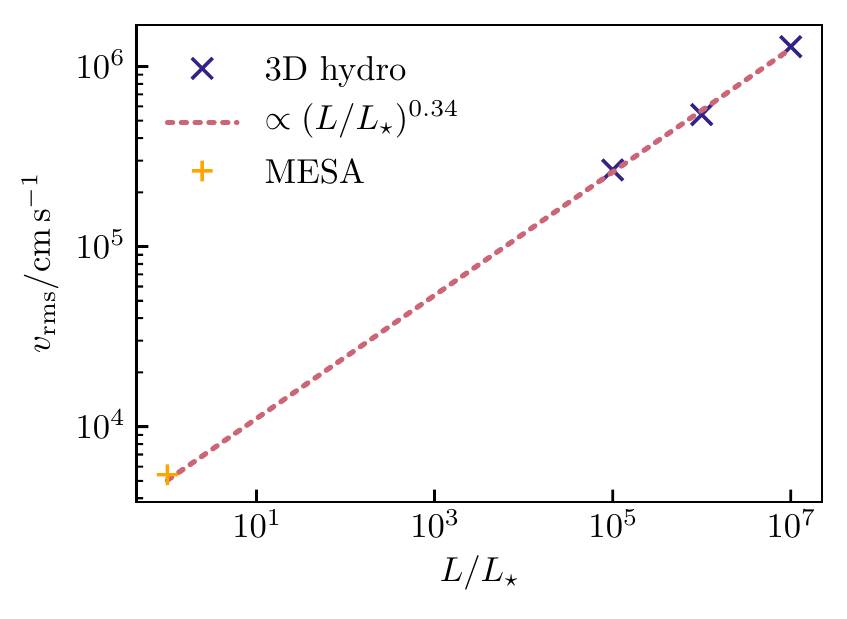}
  \caption{\label{fig:L-vconv}Relation of increased stellar luminosity and rms velocity in the convection zone. The dotted line is a power law fit to data from the 3D hydrodynamics simulations. The MESA value computed from a volume average of the MLT velocity is plotted for comparison.}
\end{figure}
As we have to increase the heating term~$\overline{Q}$ (equivalent to an increase in luminosity~$L$), thermal diffusivity~$\kappa$, and kinematic viscosity~$\nu$ for numerical reasons,
the convective velocities are higher than those predicted by mixing-length theory (MLT) using quantities from the stellar evolution model. Figure~\ref{fig:convvel-compare} compares the angular average of velocities of the different models to the MLT value. In the convection zone all our simulations have velocities one to two orders of magnitude higher than the MLT value. This causes waves at the convective--radiative boundary~(CB) to be excited at higher amplitudes, which is intended to offset the increased dissipation within the RZ with the hope of surface amplitudes being more realistic.
The rise of velocity close to the largest radii is related to the outer boundary condition.

The scaling of convective velocities with changing luminosity has been subject of previous studies. Other hydrodynamic simulations of convection zones in stars \citep{porter2000a,viallet2013a,jones2017a} find that,
\begin{equation}
  \label{eq:L-vconv}
  L \propto v_\text{rms}^3.
\end{equation}
This is also the result found using MLT \citep[e.g.,][]{kippenhahn2012a}.
The scaling relation agrees perfectly with the observations in our simulations, which fit $v_\text{rms} \propto L^{0.34}$ (see Fig.~\ref{fig:L-vconv}).

\begin{figure*}
  \centering
  \includegraphics{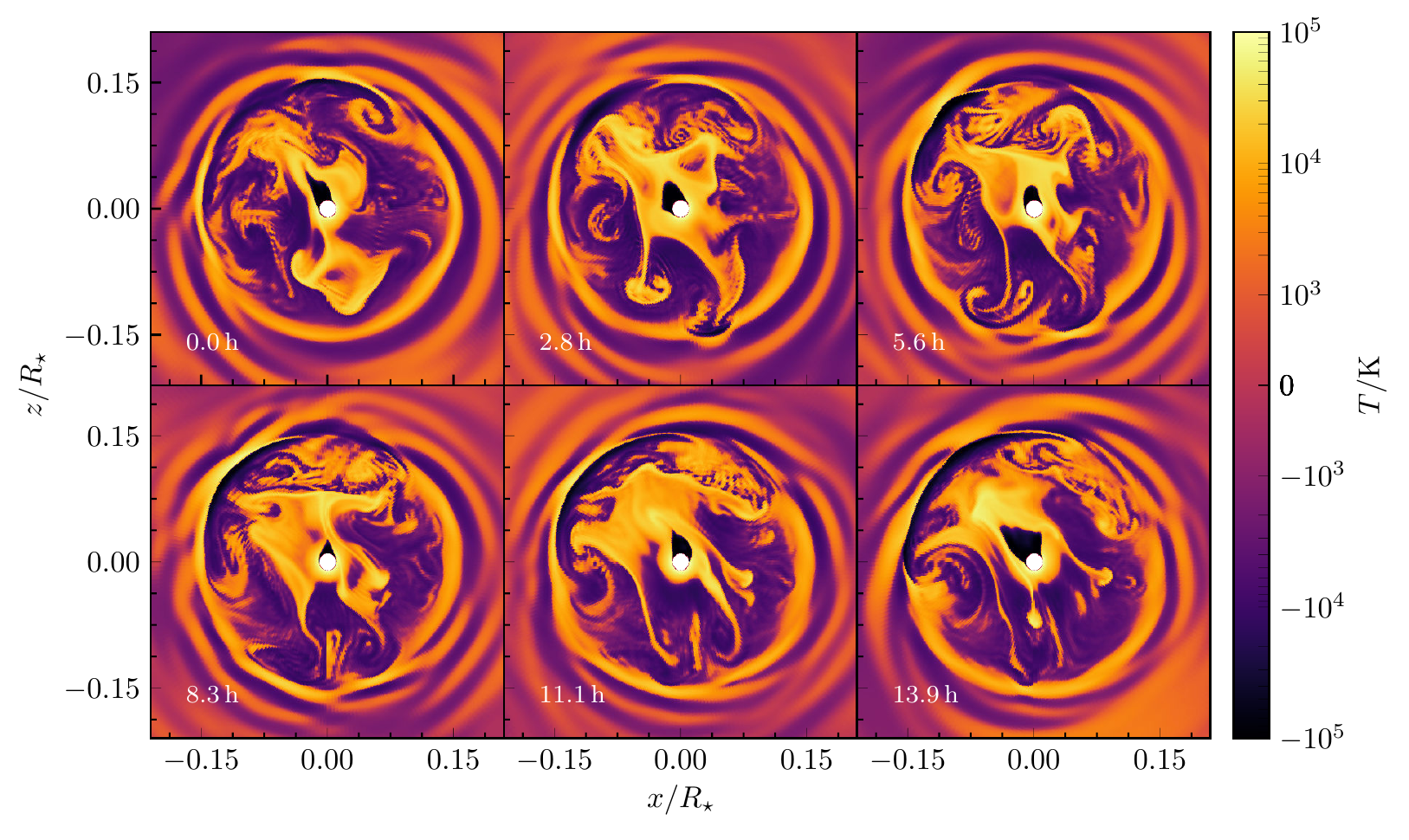}
  \caption{\label{fig:H6R10-T-series}Time series of meridional slices through model H6R10. The color scale temperature shows deviation from the horizontal mean. The label in the panel indicates time after the first panel. Each panel is separated by \num{10000} numerical time steps.}
\end{figure*}

As expected from the stratification of the 1D reference state, convection immediately starts to develop in the core. From early times convection is dominated by large plumes. These plumes often rise until they reach the CB, but are sometimes dissolved by interacting with large eddies. Their disintegration at the convective boundary perturbs the stably stratified radiation zone directly above. This process can be seen in the time series in Fig.~\ref{fig:H6R10-T-series}.

\begin{figure*}
  \includegraphics{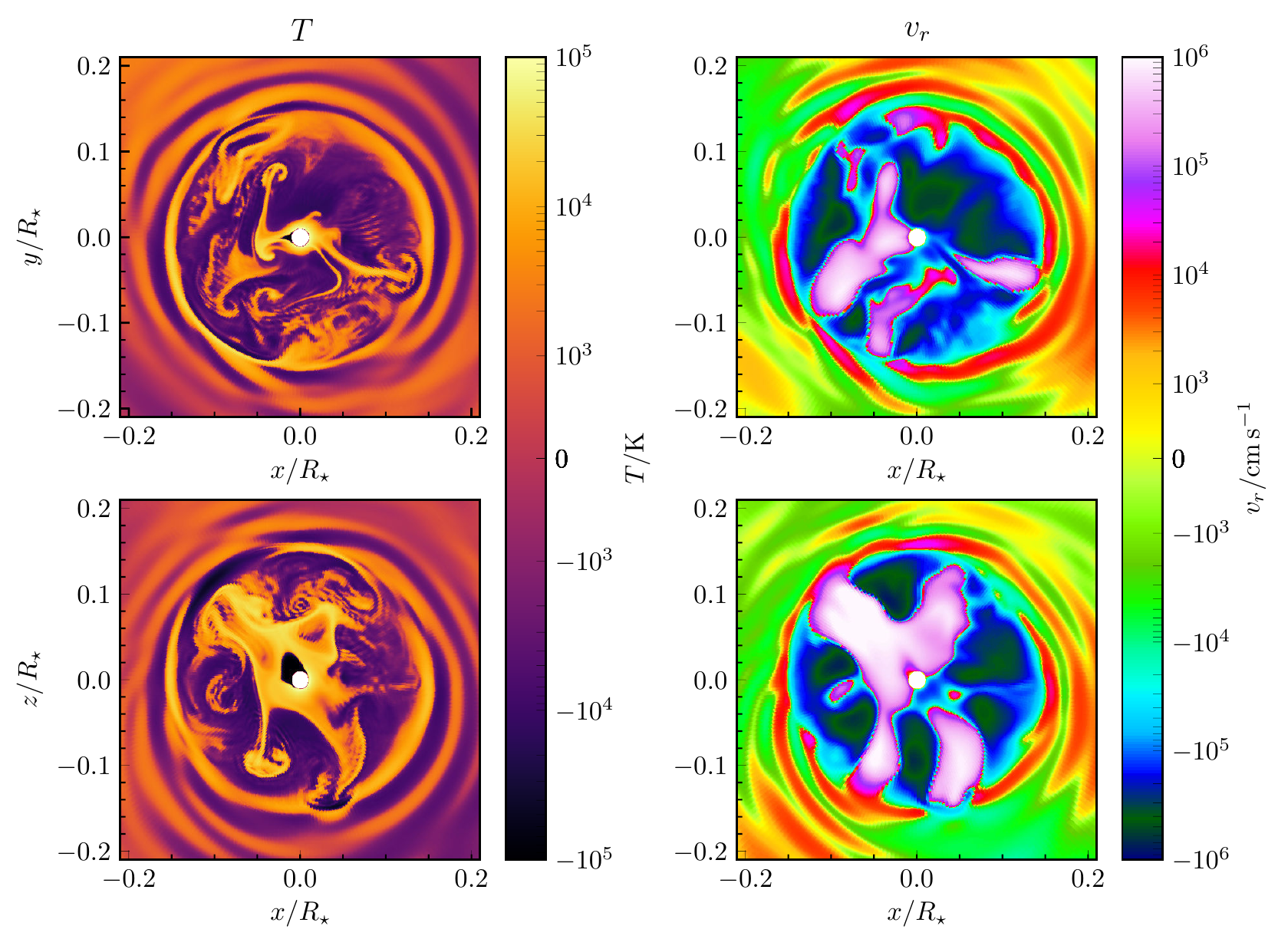}
  \caption{\label{fig:H6R10-slice}Equatorial (\emph{top row}) and meridional (\emph{bottom row}) slices through the convective core of simulation H6R10. The \emph{left column} shows the temperature deviation from the horizontal average, the \emph{right column} radial velocity.}
\end{figure*}

\begin{figure*}
  \plotone{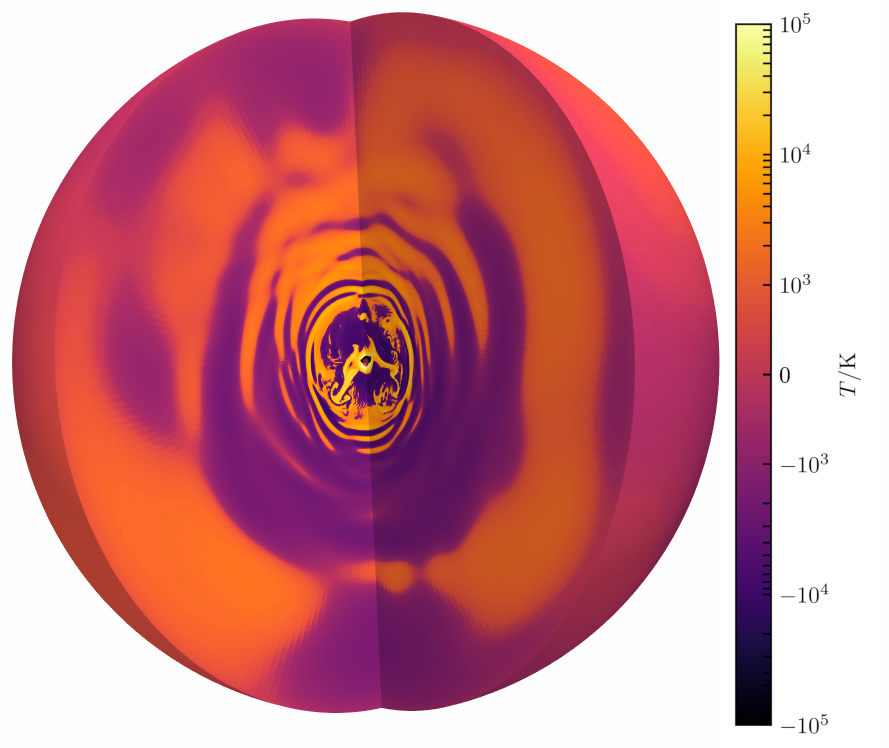}
  \caption{\label{fig:3d-temp}3D visualization of model H6R10\@. The color scale shows temperature fluctuations $T$ from the background state~$\overline{T}$. An animation of this figure is available at \url{https://www.mas.ncl.ac.uk/~npe27/videos/H6R10.html}.}
\end{figure*}

Figure~\ref{fig:H6R10-slice} illustrates the correlation of positive radial velocity and temperatures higher than the horizontal average. It shows that in model H6R10 plumes reach typical velocities of \SI{10}{km.s^{-1}}. Scaling this down to the rms convective velocities of the actual star using Eq.~\eqref{eq:L-vconv} yields a rising speed of \SI{0.1}{km.s^{-1}}. Figure~\ref{fig:3d-temp} shows a 3D view of the whole star using the same model.

The series of meridional slices in Fig.~\ref{fig:H6R10-T-series} show an example of several plumes hitting the convective boundary and triggering wave motion in the region above. Between $t=0$ and $t=\SI{2.8}{h}$ the large plume in the bottom part of the slice splits into two parts, which subsequently cause small-scale disturbances in the previously much more uniform temperature field of that region. At $t=\SI{5.6}{h}$ a larger plume hits the boundary in the upper left corner of the convection zone. It spreads out at the boundary over more than half a hemisphere and causes Kelvin--Helmholtz-like vortices on its inner side. These seem to be the cause of many of the small-scale eddies at the interface, which can themselves drive waves in the RZ.

\begin{figure*}
  \centering
  \includegraphics{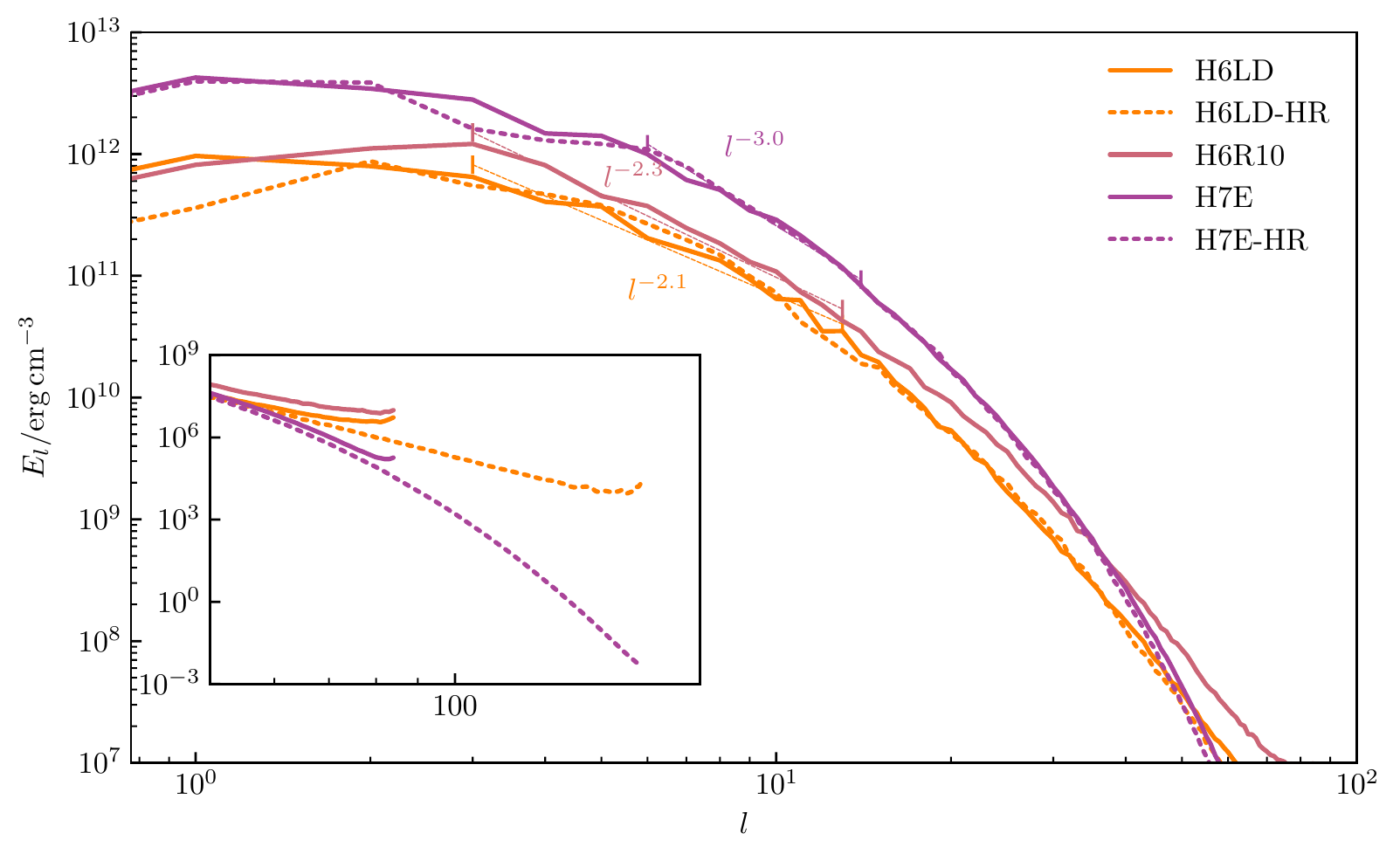}
  \caption{\label{fig:k-spectrum}Spectra of kinetic energy density in several models as a function of wavenumber. The spectra were calculated at $r=0.14\,R_\star$ as indicated in Fig.~\ref{fig:convvel-compare}. The data points for a single spherical harmonic degree~$l$ were computed by summing all azimuthal $m$ components and averaging over time. Power laws (\emph{dashed lines}) are fitted in the wavenumber ranges between the vertical markers. The inset shows the high-wavenumber tail of the high-resolution simulations H7E-HR and H6LD-HR.}
\end{figure*}

Turbulent kinetic energy in the CZ shows a typical cascade behavior, where most energy is present at low wavenumbers, i.e.\ large length scales. Figure~\ref{fig:k-spectrum} shows the kinetic energy spectrum of several models as a function of $l$ mode.
The energy contained in a single $l$~mode is computed from the poloidal ($W$) and toroidal ($Z$) decomposition (see Eq.~\eqref{eq:pol-tor-decomp}) with the expression \citep[e.g.,][Sect.~10.6.6]{glatzmaier2013a},
\begin{align}
  \label{eq:wave-energy-l}
  \nonumber
  E_l(r) &= \sideset{}{'}\sum^l_{m=0} \frac{l(l+1)}{4 \pi r^2 \overline{\rho}}\\
         & \times \left(\frac{l(l+1)}{r^2}|W_l^m|^2 + \left|\PD{W_l^m}{r}\right|^2 + |Z_l^m|^2 \right),
\end{align}
where the primed sum means that the $m=0$ terms are multiplied by $1/2$. In all cases most energy is contained in the low-order modes ($l\lesssim 5$), although the actual peak of the spectrum varies between $l=1$ and $l=3$ for the different parameters.

Although numerical diffusivity limits the inertial range in these spectra, we can still obtain a power law slope for each of the models. The slope becomes negatively steeper with increased convective forcing.
In Fig.~\ref{fig:k-spectrum} we fit the inertial range of each model with power laws.
In the strongly forced models~H6LD and H6R10, in which we see a strong influence of rising plumes (see Fig.~\ref{fig:H6R10-T-series}), the kinetic energy spectrum drops with $l^{-2.1}$ or $l^{-2.3}$, respectively. This approaches the value predicted by Bolgiano--Obukhov scaling of $l^{-2.2}$ for buoyancy-driven turbulence \citep{obukhov1959a,bolgiano1959a}. The more strongly forced models H7E and H7E-HR, show an even steeper slope in the inertial range, following $l^{-3.0}$. This is significantly steeper than the $l^{-5/3}$ relation predicted by \citet{kolmogorov1941a}, which forms the basis for theoretical spectra using the eddy excitation mechanism. This might explain why our simulations show a different slope in the frequency spectra.

The deviation from theoretically predicated slopes might be due to the relatively low Reynolds numbers reached in the simulations (see Tab.~\ref{tab:models}). On the other hand, the case of heating concentrated in a small spherical region is quite different from the plane-parallel, Boussinesq convection underlying some theoretical models and the velocity field is not necessarily isotropic in this case.

Comparing with previous hydrodynamic simulations we find that Model H6R10 agrees well with a comparable model from \citet{rogers2013a}, who find a broken power law fit with exponents of \num{-4.8} and \num{-1.9} and a break at $l\approx 10$ in a singular-value decomposition of the frequency and wavenumber spectrum. \citet{augustson2016a} obtain a qualitatively similar spectrum in their simulations of magnetic turbulence, with a low-wavenumber exponent of approximately \num{-3} and a steeper power law for higher wavenumbers. Their simulations also have a peak in the kinetic energy spectrum at low spherical harmonic degree, in the range from $l=1$ to $l=10$. We note that the models of \citet{augustson2016a} do not have enhanced forcing and yet show a similar spectrum to those in this work. This indicates that the spectrum is more dependent on the regime that nondimensional numbers like \Ra{} and \Rey{} are in than the actual value of convective forcing, as expected.

The inset in Fig.~\ref{fig:k-spectrum} shows the comparison of the high-wavenumber tail for the simulations H7E and H7E-HR, which are run with identical parameters except for the number of spectral modes being used. Their spectra are almost identical apart from a small bend at the highest $l$~values. This suggests that enough of the inertial subrange of the turbulent cascade is resolved to get the correct energy dissipation and that the simulations do not suffer from severe anomalous behavior at the smallest resolved length scales. The same is true for H6LD and its high-resolution counterpart H6LD-HR.

\begin{figure}
  \includegraphics{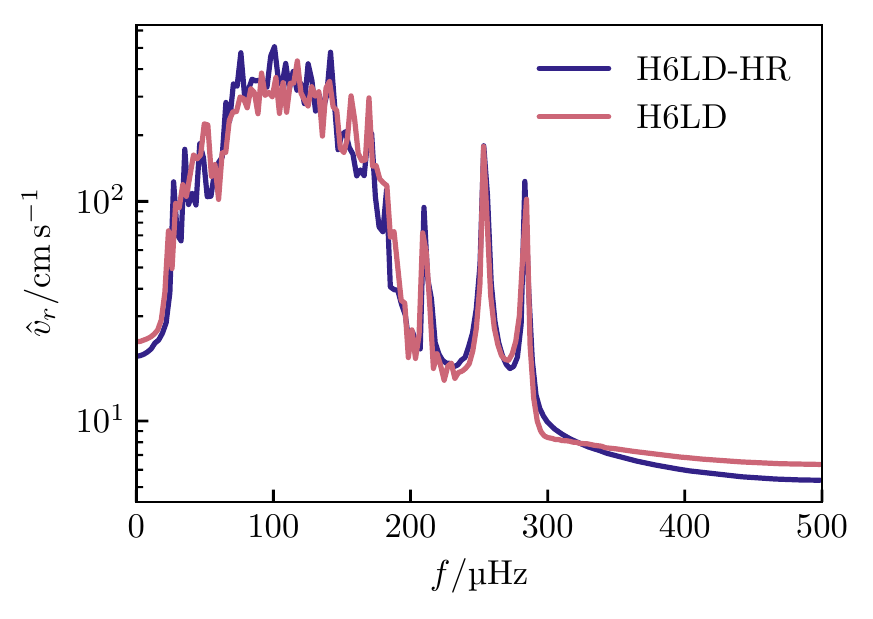}
  \caption{\label{fig:H6LD-HR-spec}Frequency spectra of radial velocity in simulations H6LD ($128\times256$ angular resolution) and H6LD-HR ($256\times512$ angular resolution). The spectrum is shown in the radiation zone at $r=0.74\,R_\star$. The number of time samples in H6LD was reduced to match H6LD-HR, which was run for a shorter time.}
\end{figure}

As a main concern of this paper are the IGW spectra, we also assess the impact of angular resolution on these. Figure~\ref{fig:H6LD-HR-spec} shows the frequency spectrum of radial velocity in the radiation zone for the two simulations H6LD and H6LD-HR, where both are identical except for latter having twice the angular resolution. We see that both simulations are very similar, including the magnitude and position of the modes between \SI{200}{\micro Hz} and \SI{300}{\micro Hz}, the continuous signal between \SI{20}{\micro Hz} and \SI{200}{\micro Hz}, and the low frequency drop due to radiative damping at \SI{20}{\micro Hz}.

\subsection{IGW generation}
\label{sec:conv-spectra}

\begin{figure}
  \includegraphics{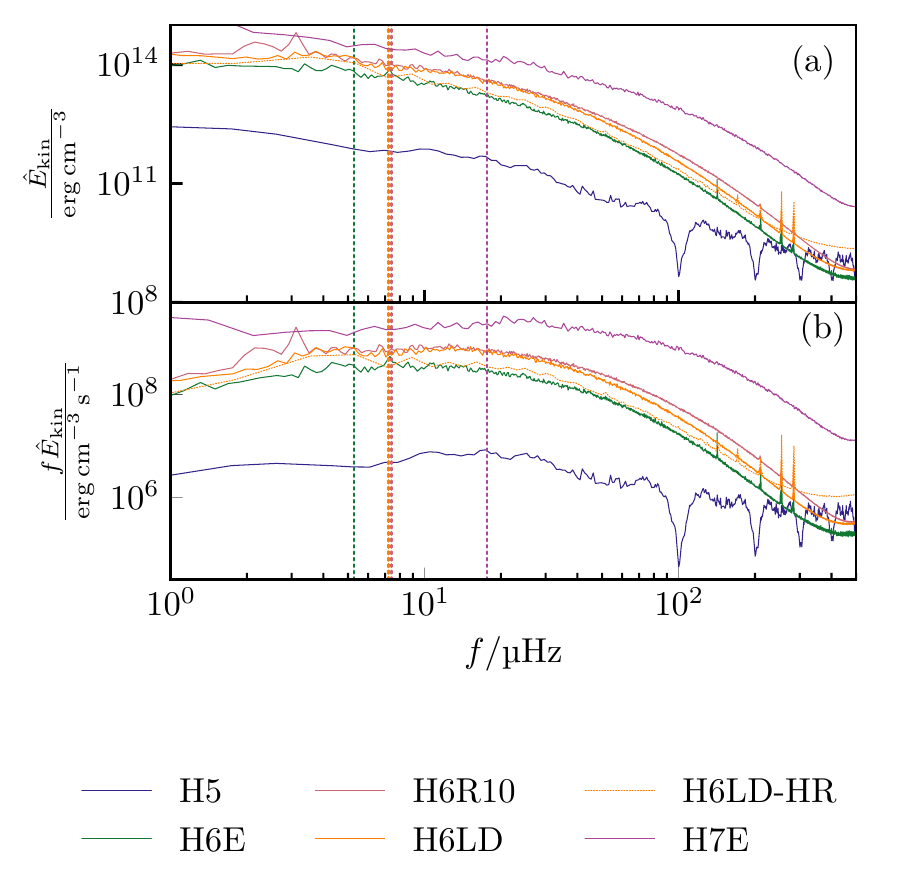}
  \caption{\label{fig:CZ-spec-all}Frequency spectra of kinetic energy just above the CZ ($r=\SI{2e10}{cm}=0.14\,R_\star$). The vertical dotted lines represent an estimate for the convective turnover frequency from Eq.~\eqref{eq:turnover}. Panel~(a) shows the Fourier transform according to Eq.~\eqref{eq:ekin-spec}. Panel~(b) is the same multiplied by frequency to account for integration over a logarithmic coordinate.}
\end{figure}

\begin{figure}
  \includegraphics{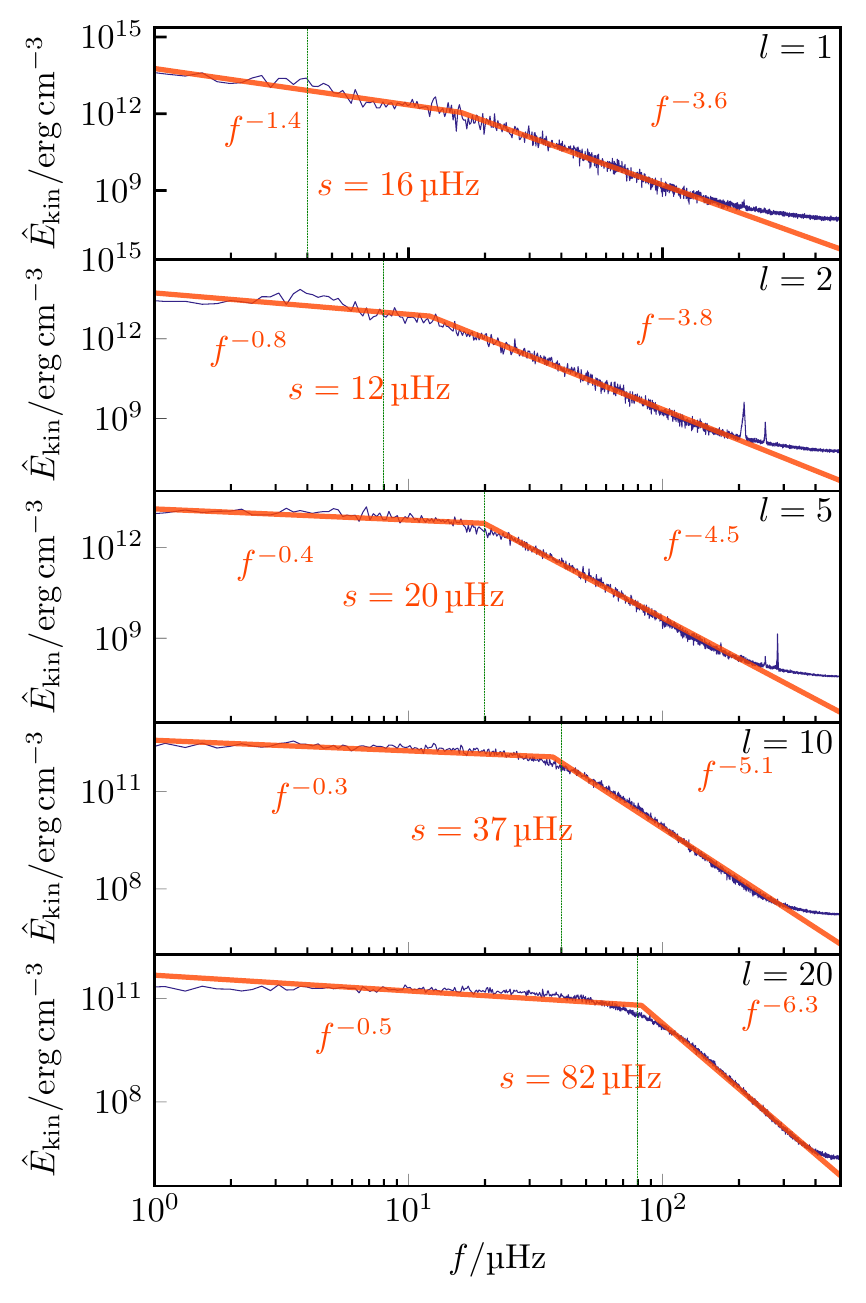}
  \caption{\label{fig:CZ-spec-l}Absolute value of the Fourier transform of kinetic energy in particular $l$~modes according to Eq.~\eqref{eq:ekin-spec}. The data are taken from simulation H6LD\@. The logarithmic data are fitted with a broken power law (\emph{orange line}). Its slopes and the position of the break are indicated next to the fit. The vertical dashed lines are estimates for the position of the break from Eq.~\eqref{eq:turnover-l}.}
\end{figure}

It is controversial which physical mechanism is most important for the excitation of IGWs at the CB\@. The two common candidates are bulk Reynolds stresses produced by convective eddies \citep{lighthill1952a,goldreich1990a} and plume overshoot \citep{townsend1966a,zahn1991a}. Most theories about the effect of IGWs in stellar interiors \citep[e.g.,][]{talon2005a,fuller2014a} employ the spectrum of IGWs derived from convective eddies \citep{kumar1999a,lecoanet2013a}. Therefore our analysis focuses on this spectrum, but the plume spectrum is considered later.

To study the spectrum of waves generated, we first investigate the spectrum of motions generated at the CB\@.
We analyze our 3D data by computing the spectrum of kinetic energy density at a radius of 0.07~$H_P$\footnote{The pressure scale height is defined as $H_P=-\partial r / \partial \ln P$.} above the top of the convection zone (as defined by the Schwarzschild criterion). This spectrum is given by,
\begin{equation}
  \label{eq:ekin-spec}
  \hat{E}_\mathrm{kin} = \frac{1}{2} \bar{\rho} \left(\hat{v}_r^2 + \hat{v}_\theta^2 + \hat{v}_\phi^2 \right),
\end{equation}
with the Fourier transforms of the individual velocity components, $\hat{v}_r$, $\hat{v}_\theta$, $\hat{v}_\phi$, according to Eq.~\eqref{eq:DFT}. Figure~\ref{fig:CZ-spec-all} shows the spectra for different models. For guidance we show an estimate of the convective turnover frequency given by,
\begin{equation}
  \label{eq:turnover}
  f_\text{TO} = \frac{v_\text{rms}}{\pi r_\text{CZ}},
\end{equation}
which assumes the largest eddy extends from the center of the star to the radius of the convection zone~$r_\text{CZ}$ and it turns at the rms velocity. Panel~(b) shows the spectra multiplied by $f$ to account for integration over $d \log f$, which makes it easier to see the regions containing most energy in the logarithmic plot. We see that, while the peak is not too far from $f_\text{TO}$, the distribution is almost flat in the low frequency regime.

Clearly, this is a spectrum of motions at this radius and is not necessarily waves (although see Sect.~\ref{sec:igw-identification}). However, this motion is what drives the waves and if it has a high-frequency component then high-frequency waves can be efficiently driven.

At this radius the integrated (i.e.\ including all harmonic degrees~$l$) frequency spectrum is nearly flat with a transition to a more steeply declining power law at higher frequencies ($f\gtrsim \SI{20}{\micro Hz}$). The spectrum is not dominated by values at $f_\text{TO}$ and indeed it is hard to make out this frequency in the spectrum. However, if we look at the frequency spectrum at particular length scales, by selecting individual values of~$l$, we start to see a sharp transition between the power laws at low and high frequency as evidenced in Fig.~\ref{fig:CZ-spec-l}.

In this scale-dependent spectrum the break point between the two power laws depends mostly linearly on angular degree~$l$ and can be approximated with,
\begin{equation}
  \label{eq:turnover-l}
  s = \SI{4.0}{\micro Hz} \cdot l.
\end{equation}
The slope is not too far from the estimate for the convective turnover frequency $f_\text{TO}=\SI{7.2}{\micro Hz}$ for this model, considering the uncertainty in the estimate of $f_\text{TO}$ in Eq.~\eqref{eq:turnover}.
This fits the conjecture by \citet{rogers2013a} that the eddy mechanism efficiently generates waves below this frequency.

It is worth noting that, even in this scale-dependent spectrum in which the break between power laws corresponds to the scale-dependent turnover frequency, the energy is \emph{not} concentrated at that frequency. This is in stark contrast to the theoretical predictions which posit that the frequency spectrum (within the CZ) is strongly peaked at the convective turnover frequency.

\begin{figure}
  \includegraphics{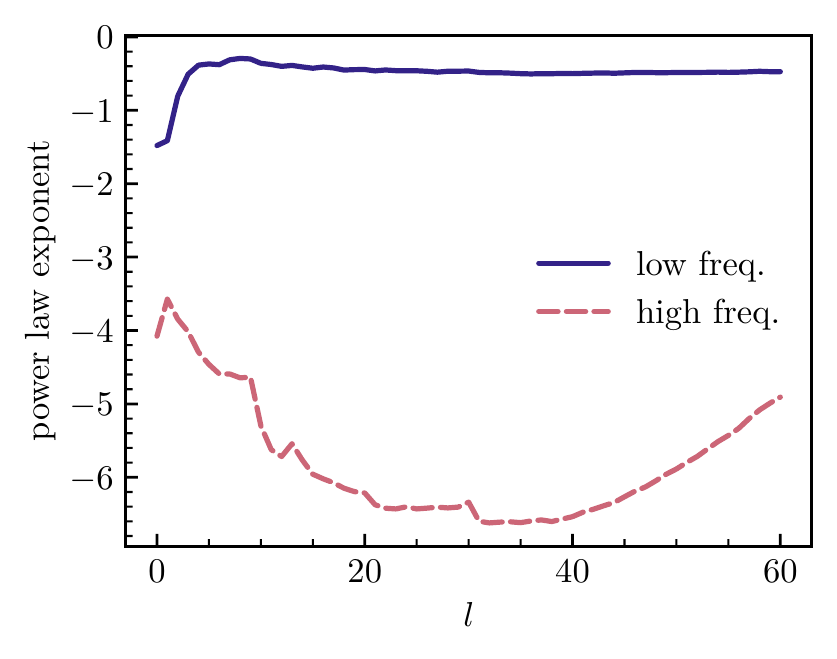}
  \caption{\label{fig:conv-slopes}Exponents of a broken power law fit to the kinetic energy spectrum on top of the convection zone of model H6E as a function of $l$~mode in the spherical harmonic decomposition.}
\end{figure}

\begin{figure}
  \includegraphics{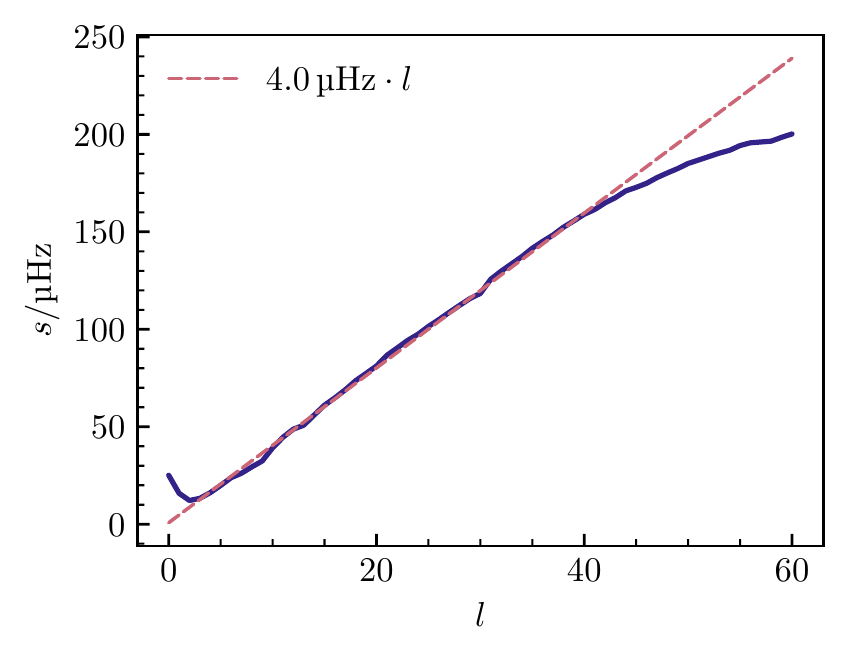}
  \caption{\label{fig:conv-break}Position of the break $s$ of the broken power law fit to the kinetic energy on top of the convection zone of model H6LD as a function of $l$~mode in the spherical harmonic decomposition. The dashed line is the turnover frequency for the particular $l$~mode, as estimated in Fig.~\ref{fig:CZ-spec-l}.}
\end{figure}

In a more systematic analysis of the broken power law fits to the frequency spectrum at the CB in Fig.~\ref{fig:conv-slopes}, we notice that the exponent of the low frequency regime stays relatively constant for $l>3$ at a value of $\num{-0.46}\pm\num{0.07}$. The high frequency component covers a wider range of exponents from \num{-3.6} at $l=2$ to \num{-6.6} at $l=33$. Both exponents show very little change at higher values of $l$. The position of the frequency break point~$s$ in the power law in Fig.~\ref{fig:conv-break} on the other hand is rising with $l$ roughly following the estimate for the convective turnover frequency from Eq.~\eqref{eq:turnover} multiplied by $l$ to account for the smaller length scales at higher spherical harmonic degree. For $l\gtrsim 31$ we observe a rise in the exponent of the high frequency range, which is due to the difficulty in fitting an increasingly smaller part of the curve. This makes the determination of $s$ less certain as well. The value of $s$ lies within the range between 12 and \SI{200}{\micro Hz} in our simulations, which is a bit higher than the range of 10 to \SI{80}{\micro Hz} in the 2D simulations of \citet{rogers2013a}. This is understandable if the position of the break really depends on $f_\text{TO}$, and in turn $v_\text{rms}$, because the 2D~simulations show a lower convective velocity.

\begin{figure}
  \includegraphics{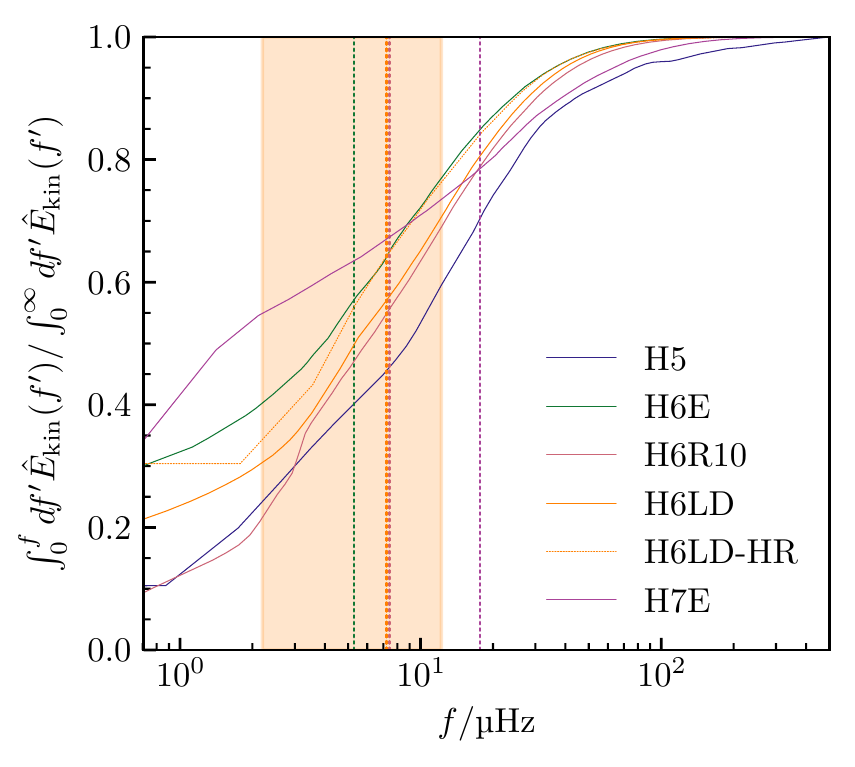}
  \caption{\label{fig:CZ-spec-integrated}Cumulative spectrum of kinetic energy density just above the CZ ($r=\SI{2e10}{cm}=0.14\,R_\star$) normalized to the value integrated over all frequencies. The vertical dotted lines represent an estimate for the convective turnover frequency from Eq.~\eqref{eq:turnover}. The shaded area shows the frequency span centered around $f_\text{TO}$ that contains 40\% of kinetic energy for model H6LD.}
\end{figure}

The logarithmic scaling of Fig.~\ref{fig:CZ-spec-all} makes it hard to see, which frequencies contribute most to kinetic energy at the top of the convection zone. To analyze this we plot the cumulative energy distribution, i.e.\ the function of energy contained below a certain frequency. It is shown normalized to the full energy of the particular model in Fig.~\ref{fig:CZ-spec-integrated}. We see that for models with a heating rate increased by a factor of $10^6$, roughly 50\% of the energy is below $f_\text{TO}$ and 50\% is above. While the steepest increase is around $f_\text{TO}$, the distribution of energy is widely spread in frequency. The shaded area shows the frequency range around $f_\text{TO}$ that contains 40\% of energy. In model~H7E with a heating rate of $10^7$ times the stellar value, 80\% of the energy is located below $f_\text{TO}$, while model~$H5$ has almost all energy far above its value of $f_\text{TO}$ at \SI{0.5}{\micro Hz}. As discussed earlier we believe model~H6LD is the best trade-off between an increased heating rate and increased diffusivity.

Comparing these numerical spectra to theoretical spectra is not straightforward as clearly the former include wave motion as well as overshooting motion (although see Sect.~\ref{sec:igw-identification}). However, one can trace differences between theoretical IGW spectra to differences in the assumed convective spectra. While the theoretical wavenumber spectrum \citep[following][] {kolmogorov1941a} has some observational basis, the frequency spectra supposed in the theoretical analysis of \citet{kumar1999a} and \citet{lecoanet2013a} -- based on the assumption of Kolmogorov scaling of eddy sizes and their corresponding turnover times -- do not. Yet it is this frequency spectrum in the CZ that determines the frequency spectra of excited IGWs.

For example, theoretical spectra do not efficiently generate high-frequency waves because of the assumption that most of the convective energy is concentrated at the convective turnover frequency. If the energy of convection itself is not limited to a narrow band around the convective turnover frequency, there is no reason to suppose that the IGW frequency spectrum would be. Moreover, if the CZ has high frequencies then it can efficiently generate waves of high frequency.
Therefore, based on comparisons of CZ spectra one can see two important issues arise between theoretical and numerical results that would affect the IGW spectra: (1) for an integrated spectrum, energy is not concentrated at the convective turnover frequency, but is spread among a wide range of frequencies; and (2) frequencies higher than $f_\mathrm{TO}$ are clearly present with significant energy within the convection zone. We also note that while Kolmogorov scaling may be the appropriate description for isotropic turbulence in a Boussinesq box, it is wholly unclear that it is appropriate for spherical configurations with a centrally peaked heating term, such as stars.

\citet{kumar1999a} mention that they deliberately ignore wave excitation by plumes due to limited information on their properties. As the flow pattern we observe in the simulations is obviously dominated by large plumes, it is natural to compare the spectra to theory of IGW excitation by plume penetration. \citet{montalban2000a} developed expressions for the IGW spectrum generated by plume penetration at the bottom of the solar convection zone. It is based on the plume model by \citet{rieutord1995a}. They explicitly caution against its use at the top of a convective envelope because of the typical importance of radiative cooling there, characterized by a low P\'eclet number ($\Pe\approx 1$). This argument does not apply here, where $\Pe \gtrsim \num{e4}$. For comparison, the stellar value in the core is $\Pe \gtrsim \num{e6}$. Therefore, for lack of a dedicated theory, we apply their model to our spectra.

The frequency dependence of the energy spectrum is determined by the plume timescale~$t_\text{b}$, which is often approximated by the ratio of plume velocity $v_\text{pl}$ and plume incursion depth $\Delta_\text{p}$. The corresponding frequency is then,
\begin{equation}
  \label{eq:plume-freq}
  f_\text{b} \sim \frac{v_\text{pl}}{\Delta_\text{p}}.
\end{equation}
The predicted energy spectrum takes the form \citep{montalban2000a},
\begin{equation}
  \label{eq:montalban-spectrum}
  E(f) \propto \exp\left(-(f / f_\text{b})^2\right),
\end{equation}
where we absorbed all factors depending on radius and wavenumber into the proportionality constant. We choose to work directly with the expression for kinetic energy instead of wave flux because the spectrum is taken right at the top of the CZ, where a conversion to flux is not straightforward.

\begin{figure}
  \includegraphics{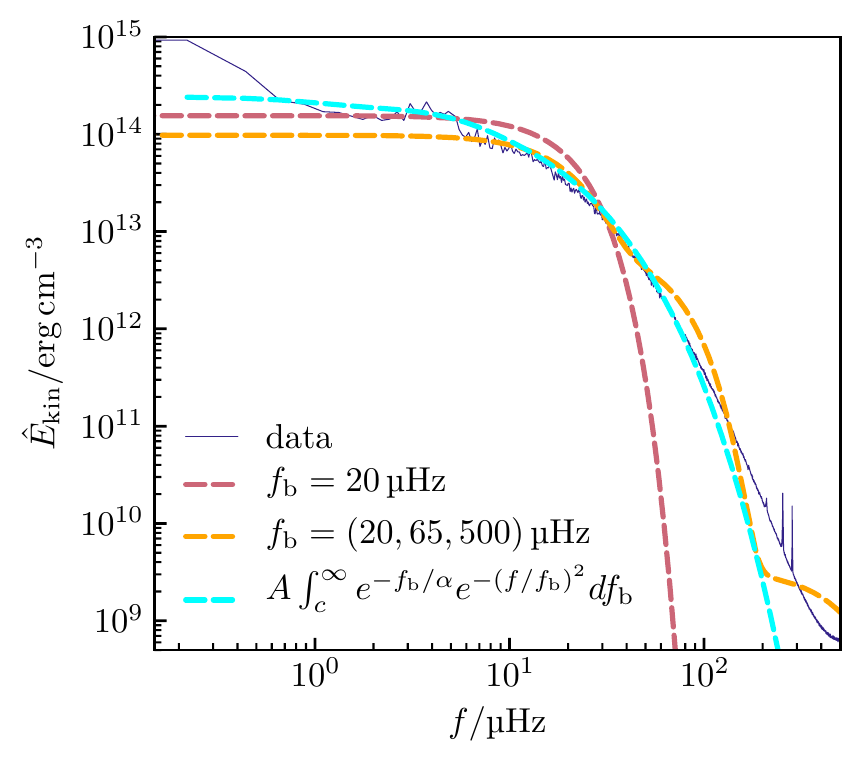}
  \caption{\label{fig:plume-spec}Kinetic energy spectrum on top of the convection zone of model~H6LD\@. The dashed lines show the theoretical spectrum for plume excitation from Eq.~\eqref{eq:montalban-spectrum}. The red line is the case of a single plume frequency~$f_\text{b}$. The orange line is a combination of three different frequencies. The cyan line is using a plume frequency distribution following Eq.~\eqref{eq:plume-fit}.}
\end{figure}

A combination of plumes with different timescales can be fit to the simulation spectra. Figure~\ref{fig:plume-spec} shows fits with one and three values of $f_\text{b}$. This shows that the plume spectrum as described by Eq.~\eqref{eq:montalban-spectrum} generally fits the shallow power law in the low-frequency regime very well, even with a single plume frequency (red line). With just three plume timescales (orange line) it is possible to fit most of the spectrum. Assuming plumes are distributed so that their frequencies follow an exponential function with a low frequency cut-off, allows us to fit the whole spectrum apart from the high frequency turnoff (cyan line).
This heuristic expression has the form,
\begin{equation}
  \label{eq:plume-fit}
  A\int_c^\infty e^{-f_\text{b}/{\alpha}} e^{-(f/f_\text{b})^2} df_\text{b}
\end{equation}
where the shape of the exponential is given by $\alpha=\SI{11.0}{\micro Hz}$ and the low-frequency cut-off is $c=\SI{1}{\micro Hz}$. It should be noted that similar fits can be obtained with other steeply declining functions (e.g., a power law with a negative exponent) for the plume frequency distribution.

\begin{figure*}
  \centering
  \includegraphics{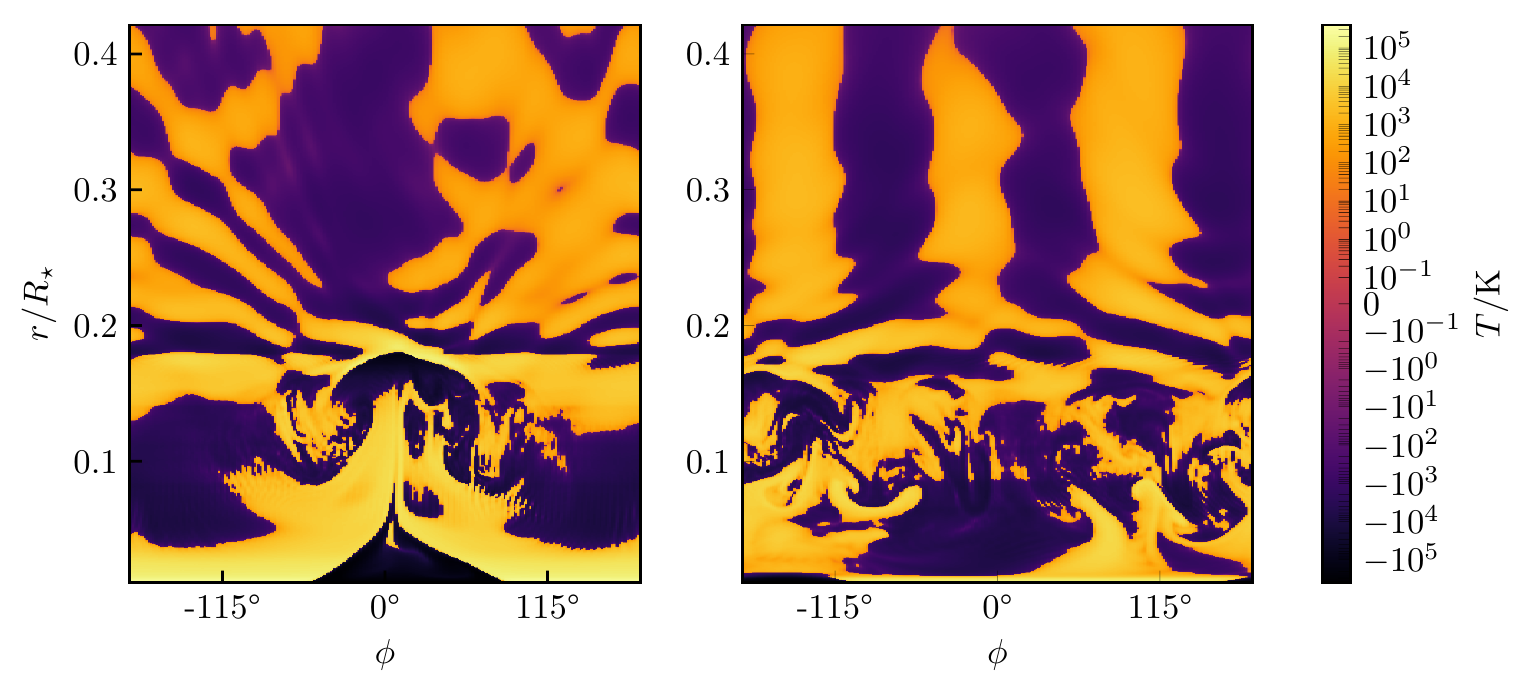}
  \caption{\label{fig:plume-rect}Two snapshots of the convective--radiative boundary (CB) in model H6LD showing the temperature deviation from the horizontal mean. The slices in radius and angle are displayed in a Cartesian projection to highlight the phase angle with respect to the CB\@. The left shows IGW excitation due to a single large plume, which results in a large angle. The right panel shows excitation by smaller eddies, which results in almost horizontal waves.}
\end{figure*}

To acquire an estimate for typical plume length and time scales from our simulations, we study the process of plume incursion in more detail. Figure~\ref{fig:plume-rect} shows a Cartesian projection of temperature perturbation at the CB in two different cases. In the left panel a large plume hits the boundary exciting waves at a large range of phase angles, including very steep angles. A representation of excitation through eddies is shown in the right panel. It results in much smaller phase angles. Both cases are remarkably similar to 2D~simulations \citep[Fig.~4]{rogers2013a}.

\begin{figure*}
  \centering
  \includegraphics{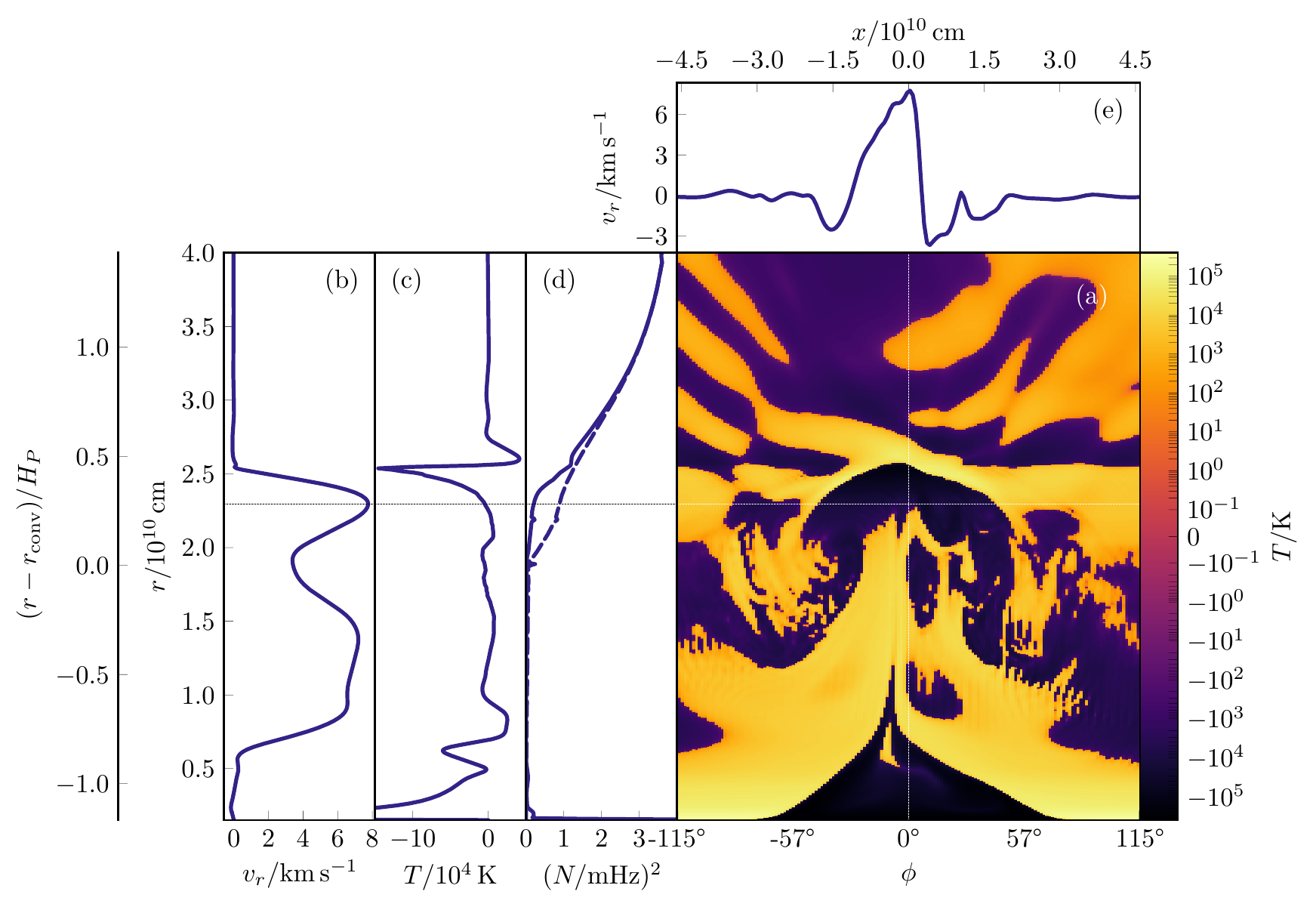}
  \caption{\label{fig:plume-detail}Close-up view of the plume in the left panel of Fig.~\ref{fig:plume-rect}. It was plotted against radius~$r$ and angle~$\phi$ to show the convective boundary as flat. The color coding in panel (a) signifies the temperature deviation~$T$ from the reference state. Panels~(b) and (c) show the vertical profile of radial velocity~$v_r$ and $T$ along the center of the plume (vertical dotted line in (a)). The solid line in Panel~(d) is the vertical profile of the square of the \bvf{}~$N^2$ computed from the current temperature profile (initial value as dashed line). Panel (e) is the horizontal profile of $v_r$ at a fixed radius (horizontal white line, position of radial maximum of $v_r$) in Panel~(a). The additional y-axis on the left shows the distance from the convective boundary in units of pressure scale height~$H_P$.}
\end{figure*}

The large plume of Fig.~\ref{fig:plume-rect} is studied in greater detail in Fig.~\ref{fig:plume-detail} to extract its size, velocity, and penetration depth. The \bvf{}~$N^2$ (Panel (d)) is significantly reduced in the overshooting region above the original convective boundary. This coincides almost perfectly with the penetration depth of the plume, which can be identified by $v_r$ approaching 0 (Panel (b)) and a discontinuity in $T$ (Panel (c)). The penetration depth is $\Delta_\text{p}=0.4\,H_P$ from the original convective boundary. The maximum plume velocity is $v_\text{pl}=\SI{7.8}{km.s^{-1}}$. This allows us to estimate a plume incursion time $t_\text{b} \sim \frac{\Delta_\text{p}}{v_\text{pl}}=\SI{2.2}{h}$, which corresponds to a frequency~$f_\text{b}$ of \SI{128}{\micro Hz}. The lateral extent of the plume~$b$ can be defined as the region of positive~$v_r$ in Panel (e). It has a value of $b=\SI{1.4e10}{cm}$ in this case. This is also sometimes used to compute the plume timescale, which results in a value of \SI{56}{\micro Hz} here. As it is easier to extract from simulations in a systematic way, we stay with the first definition using $\Delta_\text{p}$ in the following analysis.

\begin{figure}
  \centering
  \includegraphics{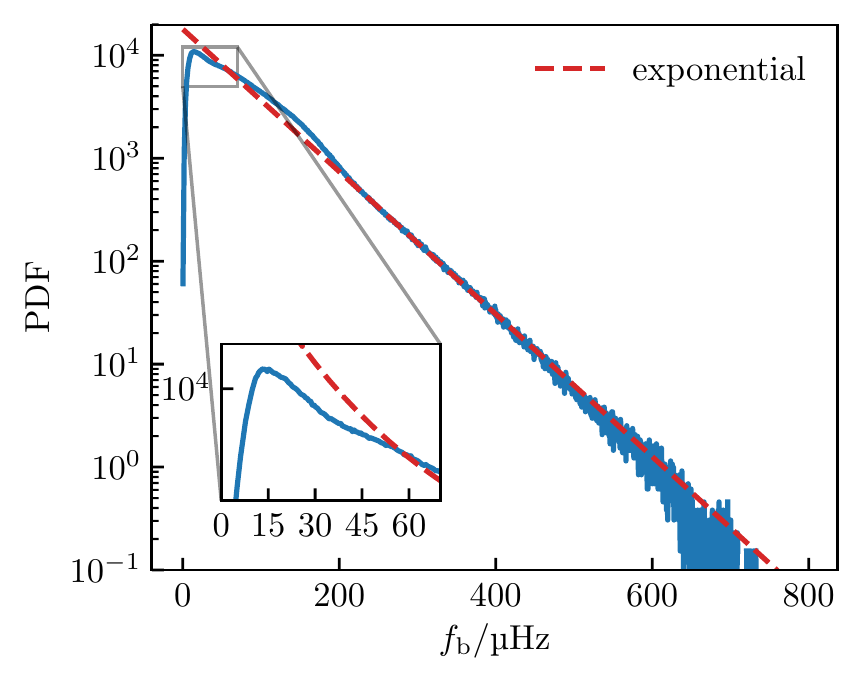}
  \caption{\label{fig:plume-freq-pdf}Probability density function (PDF) of plume frequencies. It was extracted from simulation H6LD and computed using Eq.~\eqref{eq:plume-freq}. The red, dashed line is an exponential fit to the high-frequency ($f_\text{b}>\SI{210}{\micro Hz}$) end of the PDF. The inset shows a zoom on the peak of the PDF.}
\end{figure}

We apply this estimate of the plume frequency statistically to all plumes in simulation H6LD. For each longitude and latitude, and each output snapshot (every \SI{1000}{s}) we determine if there is a rising plume and in that case compute a plume incursion depth and plume velocity. The criterion for a plume is that $v_r$ is positive at the position of the convective boundary. The incursion radius~$r_\text{p}$ is then defined as the radius at which $v_r$ first becomes negative along a line at this particular angle. The penetration depth is calculated as the distance to the convective boundary $\Delta_\text{p}=r_\text{p}-r_\text{conv}$. The plume velocity~$v_\text{pl}$ is the highest value of $v_r$ between the top of the convective boundary and $r_p$. Using the estimate for the plume frequency from Eq.~\eqref{eq:plume-freq}, we compute the probability density function (PDF) of $f_\text{b}$ throughout the simulation (Fig.~\ref{fig:plume-freq-pdf}). It rises sharply to its maximum at \SI{15}{\micro Hz} and then drops roughly following an exponential distribution.
An exponential fit to the data does not perfectly match the values found when fitting Eq.~\eqref{eq:plume-fit}. The parameter~$\alpha$ is too high by a factor of 3. Yet considering the simplistic definition of $f_\text{b}$, this still makes a strong argument for an exponential distribution of plume frequencies as the explanation of a large part of the kinetic energy spectrum at the top of the convection zone and hence, the IGW frequency spectrum.

To understand the effect of increased forcing and diffusivity we follow the discussion of plume lifetimes of \citet{pincon2016a}. They argue that plume velocity scales with luminosity as $v_\text{pl}\propto L^{1/3}$ which is consistent with the scaling of the convective velocities from Eq.~\ref{eq:L-vconv}. A luminosity increased by a factor of $10^6$ would thus result in $v_\text{pl}$ increased by a factor of 100. The penetration depth~$\Delta_\text{p}$ is not expected to be strongly affected by the change in forcing (see end of Sect.~\ref{sec:overshoot} for an estimate). The effect of radiative thermalization, while strongly increased due to the higher value of $\kappa$ in the simulations, is still negligible as the timescale~$t_\text{rad}\sim \Delta_\text{p}/\kappa$ is of the order of two years, much longer than any observed plume lifetime. The turbulent timescale inside the plume $t_\text{turb} \sim b / v_\text{pl}$, with the lateral plume size~$b$. Assuming that $b$ is not strongly affected by increased forcing, similar to $\Delta_\text{p}$, this means that plume frequency~$f_\text{b}=1/t_\text{b}$ scales like $v_\text{pl}\propto L^{1/3}$. In model H6LD ($L = 10^6 L_\star$) this results in $f_\text{b}$ being too high by a factor of 100.

\subsection{Convective overshoot}
\label{sec:overshoot}

The treatment of convective--radiative boundaries (CB) in 1D stellar evolution codes is a long-standing problem. It can have a significant impact on the evolution and nucleosynthetic signature of stars by mixing of species beyond convective regions. Hydrodynamic simulations in two or three dimensions promise insight based on first principles and have been subject of previous work \citep[e.g.,][]{freytag1996a,rogers2006a,meakin2007a,jones2017a,cristini2017a}. There is no single accepted definition of the overshooting depth in terms of angular averages of 3D~quantities. For better comparability between different stellar parameters the overshooting depth is usually stated in multiples of the pressure scale height~$H_P$ above the convective boundary as defined by the Schwarzschild or Ledoux criterion. Both criteria are equivalent in the case studied here because the star is chemically homogeneous.

\begin{figure}
  \centering
  \includegraphics{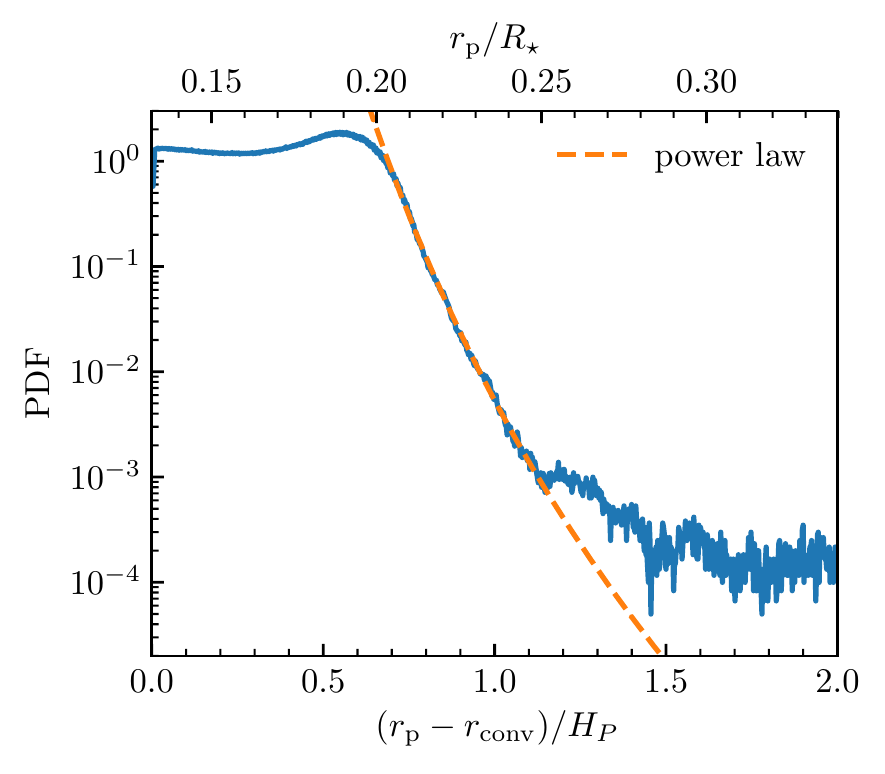}
  \caption{\label{fig:depth-pdf}Probability density function (PDF) of plume incursion depth computed from simulation H6LD\@. The distribution peaks at $0.54\,H_P$ above the convective boundary. The region after the peak was fitted with a power low with an exponent of \num{-1.4}.}
\end{figure}

The statistics of velocities and penetration depth~$\Delta_\text{p}$ from Sect.~\ref{sec:conv-spectra} can also be used to make statements on the size of the overshooting region. Figure~\ref{fig:depth-pdf} shows the PDF of plume penetration depth in model~H6LD\@. The distribution at low $\Delta_\text{p}$ is relatively flat until it peaks at $0.54\,H_P$. Beyond that it drops following a power law with exponent \num{-1.4}. This is consistent with the picture in Panel (d) of Fig.~\ref{fig:plume-detail}, where $N^2$ is affected by penetration up to a value of approximately $0.5\,H_P$. 95\% of plumes penetrate no further than $0.695\,H_P$, which is the value we will use as the boundary of the overshooting region in Sect.~\ref{sec:igw-identification}.

\begin{figure}
  \centering
  \includegraphics{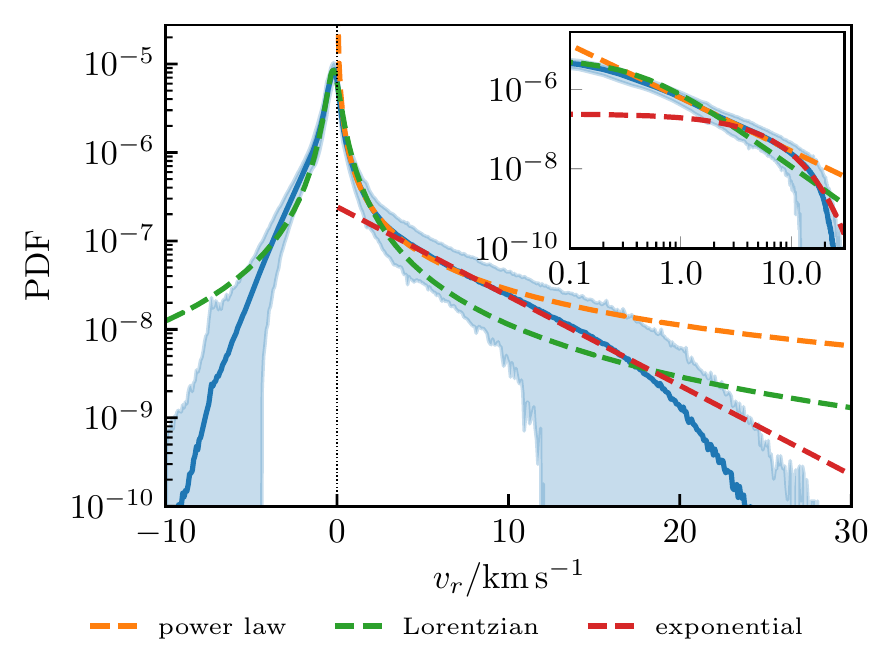}
  \caption{\label{fig:vr-pdf}Probability density function (PDF) of $v_r$ $0.42\,H_P$ above the top of the convection zone in model~H6LD (blue line). The blue shaded area signifies standard deviation over all time steps. A Lorentzian (\emph{green line}) was fitted to the central part of the distribution. A power law (\emph{yellow line}) and an exponential function (\emph{red line}) were fitted in the regions of positive $v_r$. The inset plot shows a log-log plot of the same data.}
\end{figure}

Figure~\ref{fig:vr-pdf} shows the distribution of updrafts and downdrafts in the overshooting region ($0.42\,H_P$ above $r_\text{conv}$). The PDF is peaked in Lorentzian shape at \SI{-0.2}{km.s^{-1}}. The inward velocities are distributed in a smaller range, \SI{-10}{km.s^{-1}} at most, than the outward velocities, which extend up to \SI{25}{km.s^{-1}}.

Our use of an increased convective forcing and thermal diffusivity raises the question of the validity of these results for the actual stellar values. In his study of convective penetration in stellar interiors \citet{zahn1991a} found a simple scaling law for the size of the penetrative region \citep[see also,][]{rogers2006a},
\begin{equation}
  \label{eq:zahn-Lp}
  \Delta_\text{p}^2 = \frac{3}{5} H_P H_\kappa f \frac{\rho v_\text{pl}^3}{F_\text{tot}},
\end{equation}
with the scale height of thermal diffusivity $H_\kappa = - d \ln r / d \ln \kappa$, plume filling factor~$f$, and total energy flux~$F_\text{tot}$. As $\kappa$ is only multiplied by a constant in the radiation zone, $H_\kappa$ is identical to the stellar value. The same is true for $H_P$ and $\rho$. While the simulations have an increased $F_\text{tot}$, we found the scaling $F_\text{tot} \propto v^3$ (Eq.~\eqref{eq:L-vconv} and Fig.~\ref{fig:L-vconv}). This means the penetration depth in the simulations and in the star only vary by a factor of $\sqrt{f_\text{sim}/f_\star}$, which we expect to be a number not too far from unity.

\subsection{IGW propagation}
\label{sec:radiation-zone}
\begin{figure}
  \includegraphics{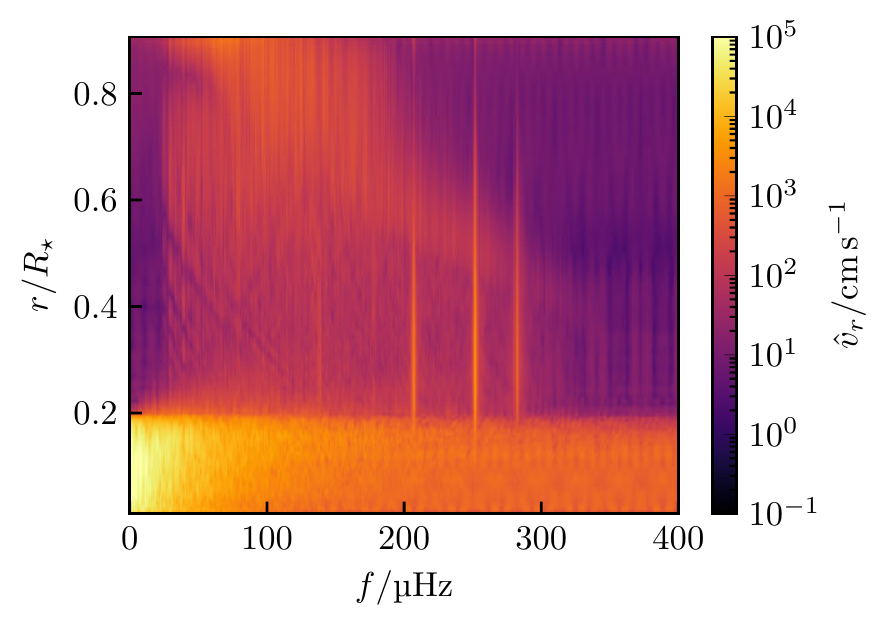}
  \caption{\label{fig:vr-spec-pcolor-full}Frequency spectrum of radial velocity at the equator of model H6LD for all radii. The values were computed by sampling 8 points at different longitudes and averaging over the absolute value of the Fourier transform.}
\end{figure}

\begin{figure}
  \centering
  \includegraphics{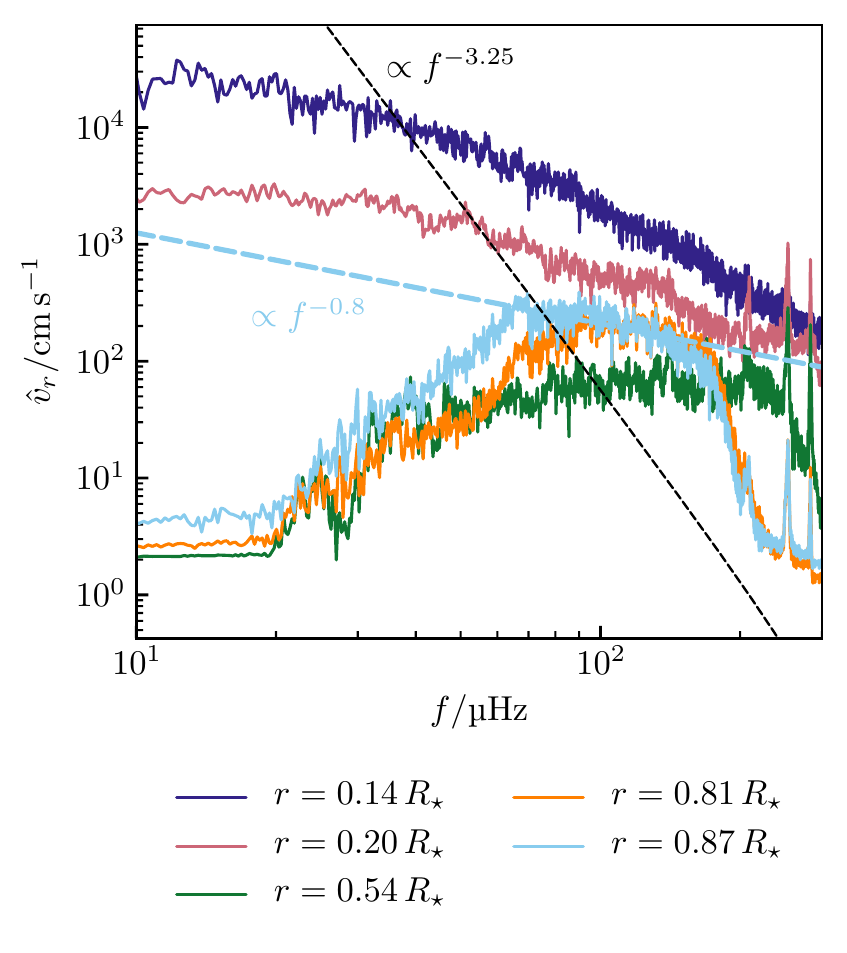}
  \caption{\label{fig:H6LD-spec-rads}Frequency spectrum of $v_r$ in simulation H6LD at different radii integrated over all~$l$. These are line plots of the spectra shown in Fig.~\ref{fig:vr-spec-pcolor-full} at several radii. The black dashed line is a theoretical prediction for the frequency dependence of $v_r$ from \cite{lecoanet2013a}. The light blue dashed line is a power-law fit to the simulation data in the range \SIrange{60}{150}{\micro Hz} at $r=0.87\,R_\star$.}
\end{figure}

\begin{figure*}
  \centering
  \includegraphics{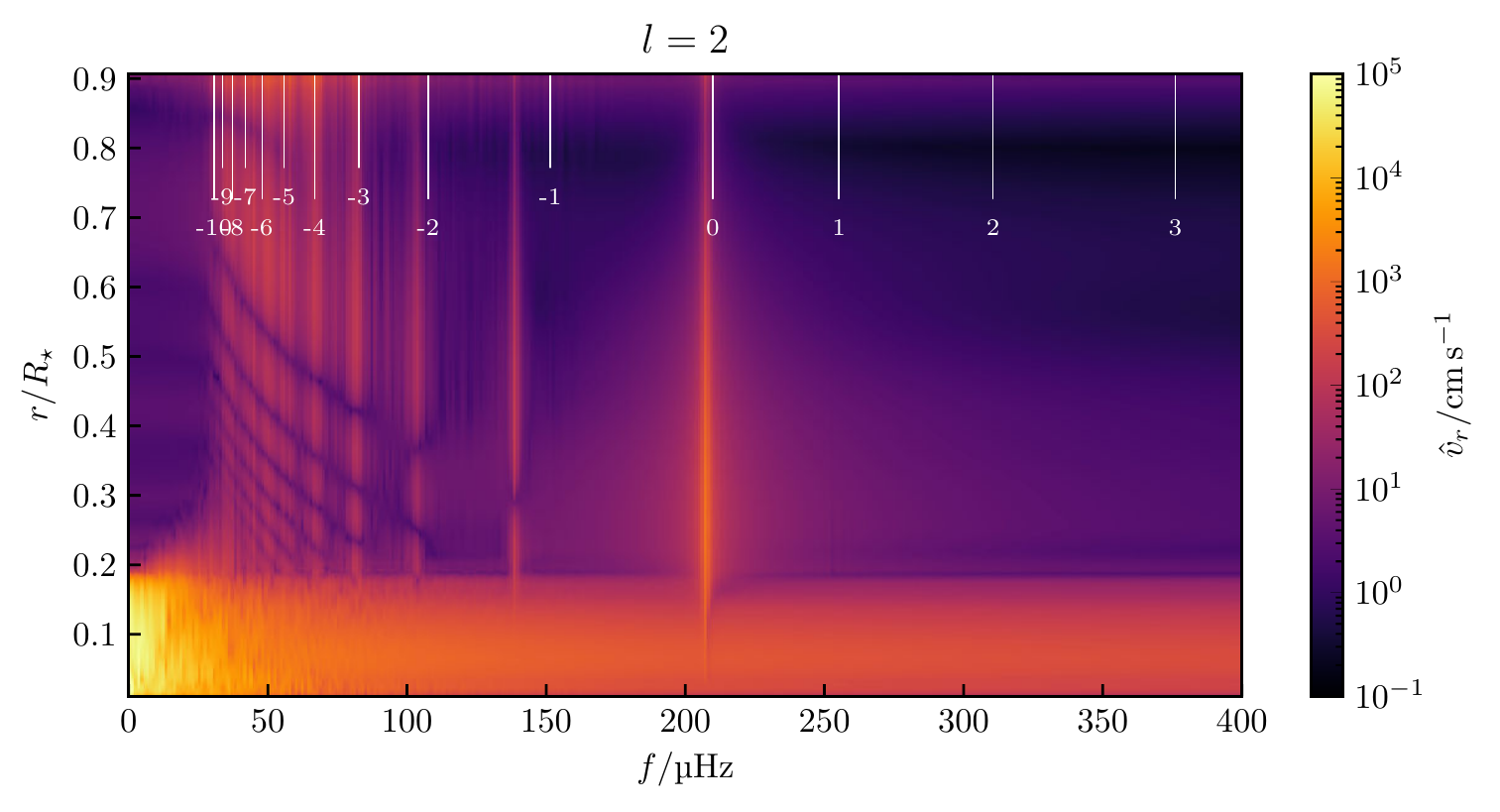}\\
  \includegraphics{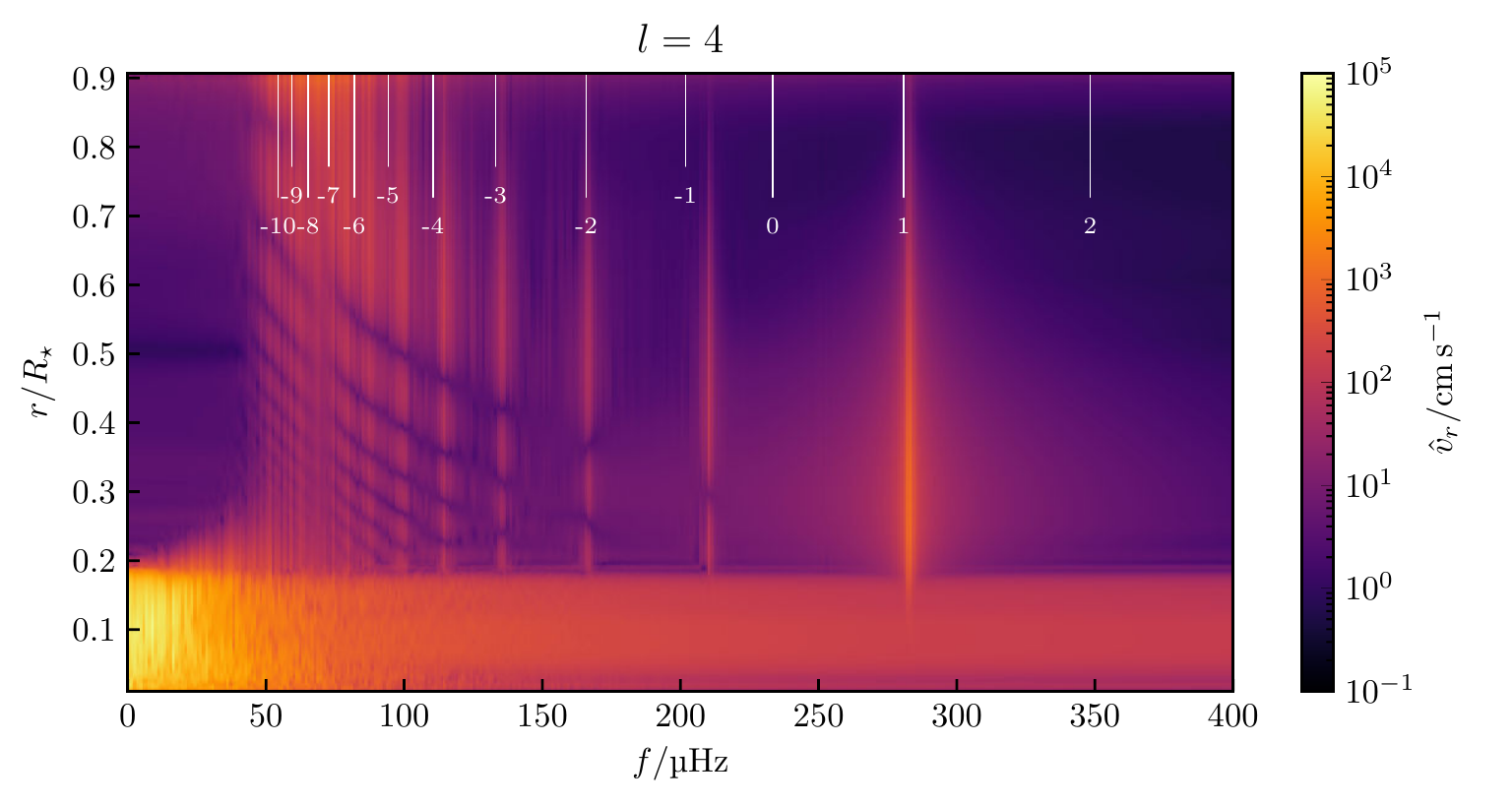}
  \caption{\label{fig:vr-spec-pcolor-l-2-4}Frequency spectra of radial velocity~$v_r$ for all radii. The $v_r$ values were sampled at several points on the equator and computed only for angular degree $l=2$ (top panel) and $l=4$ (bottom panel). The data were extracted from model H6LD\@. The vertical lines at the top are the expected mode frequencies computed with GYRE\@. The modes are numbered using the Eckart-Osaki-Scuflaire-Takata scheme, where positive numbers are p~modes, negative numbers are g~modes, and the f~mode is identified by zero. The length of the lines is varied purely for better readability.}
\end{figure*}

\begin{figure}
  \includegraphics{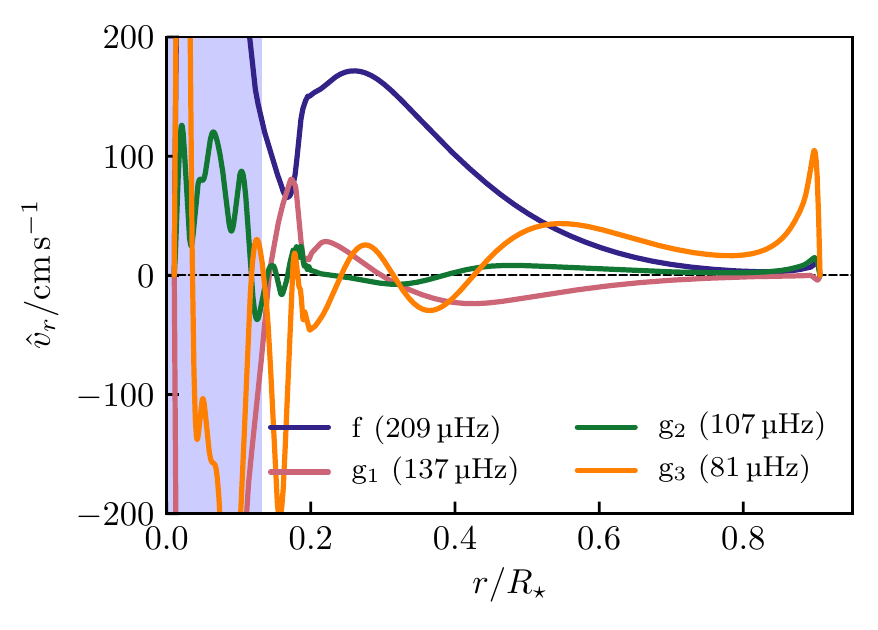}
  \caption{\label{fig:amp-l-2}Amplitude variation for different standing modes ($l=2$) from model H6LD computed for a single point at the equator. This corresponds to a vertical slice through the top panel of Fig.~\ref{fig:vr-spec-pcolor-l-2-4}. The blue shaded area is the convection zone according to the Schwarzschild criterion. The modes where identified by the number of nodes in the RZ.}
\end{figure}

Convective motions in the core generate IGWs at the CB, which propagate through the cavity of positive $N^2$ in the radiation zone. To visualize the excited frequencies and the change of the wave spectrum with radius, we compute the frequency spectrum of $v_r$ sampled at several longitudes around the stellar equator at all times for all radii
and show it as a heat map in Fig.~\ref{fig:vr-spec-pcolor-full} for model H6LD\@. In the convection zone ($r\lesssim \SI{2e10}{cm}=0.14\,R_\star$) we note the presence of all frequencies with a clear dominance of the range below \SI{50}{\micro Hz}. This is reflective of the large range of timescales of convective motion (see Sect.~\ref{sec:conv-spectra}).
In the radiation zone frequencies up to \SI{300}{\micro Hz} are excited. Low-frequency waves are strongly damped and only frequencies above \SI{40}{\micro Hz} reach the top of the simulation domain. This is qualitatively in agreement with linear theory, which predicts that lower frequency waves experience stronger damping \citep[e.g.,][]{kumar1999a}. The exact position of this lower cut-off depends on the value of $\kappa$ as well, which is why the present simulations cannot predict it quantitatively. Figure~\ref{fig:H6LD-spec-rads} shows line plots of the same spectrum at different radii. It shows that the spectrum in the RZ at low frequencies ($\lesssim \SI{20}{\micro Hz}$) does not reach the numerical noise level, as would be expected by the excessive numerical diffusion at this frequency, but turns flat at a higher value. The figure also indicates the expected frequency dependence of $v_r$  from theoretical work by \citet{lecoanet2013a}, which is $f^{-3.25}$ for the radial velocity of waves excited at a discontinuous $N$ profile. We see that the simulated spectrum is much flatter than this prediction, following $f^{0.8}$. The steep drop around \SI{200}{\micro Hz} for $r>0.8\,R_\star$ is due to the limit imposed by the \bvf{} at these radii.

Strong vertical features are visible in Fig.~\ref{fig:vr-spec-pcolor-full}. These are peaks in the spectrum which are present at the same frequency at all radii in the radiation zone. This identifies them as standing waves. Their frequencies are determined by the cavity they resonate in and can be computed numerically using the stellar oscillation code GYRE \citep{townsend2013a}.

It is hard to disentangle individual modes because the contributions of several wave numbers overlap, but due to the horizontal discretization of the simulations using spherical harmonics it is simple to extract a spectrum for particular $l$ and $m$ modes. The panels in Fig.~\ref{fig:vr-spec-pcolor-l-2-4} show the frequency spectrum for the modes $l=2$ and $l=4$. Here, the radial order of the individual standing modes can be clearly identified by the number of radial nodes. The strong mode at \SI{210}{\micro Hz} without any nodes is a \emph{fundamental mode} or f~mode. The other visible modes show an increasing number of nodes with decreasing frequency. This identifies them as g~modes \citep[e.g.,][Sect.~3.5]{aerts2010a}. We computed expected mode frequencies with GYRE\footnote{We used version 5.1 from the GYRE web page.} for comparison. They are labeled in the figure using the Eckart-Osaki-Scuflaire-Takata scheme \citep[e.g.,][]{aerts2010a}, where negative numbers indicate g~modes, 0 is the f~mode, and positive numbers are p~modes. In the case of $l=2$ we find quite good agreement for the g~modes (at least up to $\text{g}_4$) and the f~mode, especially considering that our 3D~simulation has a slightly different resonant cavity due to the different equation of state and outer boundary compared to the 1D MESA model. As expected there are no p~modes as the chosen set of equations (see Sect.~\ref{sec:method}) does not include the physics of sound waves. For $l=4$ (lower panel of Fig.~\ref{fig:vr-spec-pcolor-l-2-4}) the $\text{g}_1$, $\text{g}_2$, and $\text{g}_3$ modes match very well, while the identified f~mode is within a few \si{\micro Hz} of the expected frequency of the $\text{p}_1$~mode according to GYRE\@. This is probably coincidental as the discrepancy between 3D~hydrodynamics and GYRE gets even larger at higher wave numbers. The identification of modes by counting the number of nodes in the RZ is illustrated in Figure~\ref{fig:amp-l-2}, which shows the radial change of $v_r$ amplitude of particular frequency components corresponding to the standing waves. The amplitudes were computed by projection on a complex phase angle of the Fourier transform at the radius with the maximum absolute value in the RZ\@. The $\mathrm{g_2}$ and $\mathrm{g_3}$ show nodes at the top of the overshooting region, which is ignored for the mode identification.

\begin{figure}
  \includegraphics{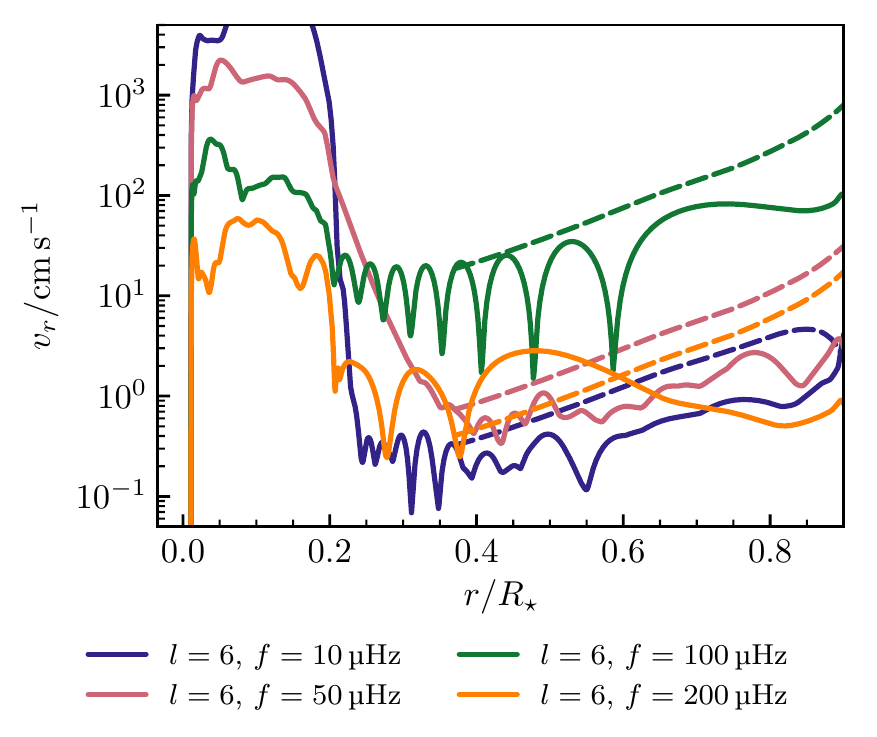}
  \caption{\label{fig:waveamp-l}Wave amplitude in $v_r$ for the $l=6$ mode at three different frequencies from simulation H6LD\@. The dashed lines are the theoretical prediction using radiative damping and pseudomomentum conservation from Eq.~\eqref{eq:wave-ampl}. This curve uses the enhanced values of thermal diffusivity~$\kappa$ from the simulation instead of the stellar values.}
\end{figure}

Linear theory predicts amplification of waves moving along a decreasing density profile through pseudomomentum conservation \citep[e.g.,][]{buehler2009a}. At the same time thermal diffusivity damps the wave. \citet{ratnasingam2019a} give an expression for the linear wave amplitude based on \citet{press1981a} and \citet{kumar1999a}. The amplitude of the radial velocity follows
\begin{equation}
  \label{eq:wave-ampl}
  v_r \propto \left(\frac{r_0}{r}\right)^{3/2} \sqrt{\frac{\rho_0}{\rho}} \left(\frac{N^2-\omega^2}{N_0^2 - \omega^2}\right)^{1/4} \exp(-\tau/2),
\end{equation}
with
\begin{equation}
  \tau = \int_{r_0}^r dr \frac{\kappa \left[l(l+1)\right]^{3/2} N^3}{r^3 \omega^4} \sqrt{1-\frac{\omega^2}{N^2}},
\end{equation}
using $\omega = 2 \pi f$ and the starting radius of wave propagation~$r_0$ with its corresponding density~$\rho_0$ and \bvf~$N_0^2$. We extract the amplitude of $v_r$ at several frequencies for a particular $l$ mode and show it together with the theoretical prediction from Eq.~\eqref{eq:wave-ampl} in Fig.~\ref{fig:waveamp-l}. We see that the waves generally follow amplification through the $\sqrt{\rho_0/\rho}$ term and are hardly affected by radiative damping, except for the low frequency case, as expected.

Generally the match between the GYRE predictions and data extracted from 3D hydrodynamics is quite promising, considering that both approaches make slightly different assumptions about the physics. Even with the high thermal diffusivity needed for the simulation we can see wave amplification. We might be able to observe wave breaking in future simulations which extend to regions closer to the surface at much lower densities.

\subsection{Nature of the signal in the radiation zone}
\label{sec:igw-identification}
\begin{figure}
  \centering
  \includegraphics{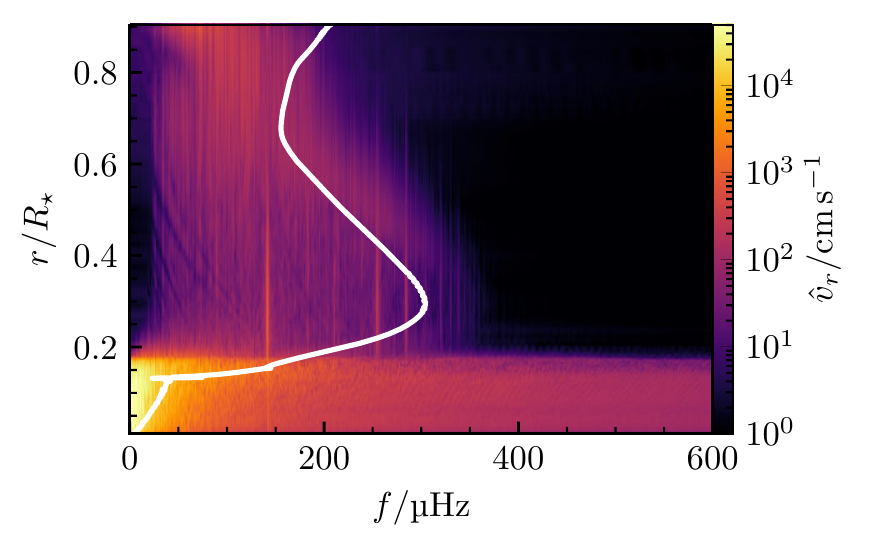}
  \caption{\label{fig:wave-brunt}High-frequency regime of the spectrum of $v_r$ in simulation H6E at all radii integrated over all $l$ components. The white line is the \bvf{}~$N/2\pi$. We see that the signal in the radiation zone is approximately limited to the region where $2\pi f<N$, with the notable exception of f~modes (strong vertical features) going beyond that limit.}
\end{figure}

Although a visualization of the temperature field in the radiation zone such as in Fig.~\ref{fig:3d-temp} suggests a wave nature of the flow field, a more rigorous analysis is needed to prove the motions are indeed IGWs excited close to the convective boundary or by nonlinear interaction in the RZ\@. IGWs are naturally limited to frequencies below the \bvf{}, i.e. $\omega < N$ with $\omega=2\pi f$. In Fig.~\ref{fig:wave-brunt} we show the high-frequency part of the spectrum of $v_r$ at all radii. The white line in the figure indicates the local \bvf{}. We see that the bulk of the signal in the RZ is constrained to the region $\omega < N$. Beyond this frequency there is a sharp drop in the amplitude which is consistent with IGW nature. A notable exception are the strong f~modes (e.g.\ at \SIlist[list-units=single]{320;330;340}{\micro Hz}) going beyond that limit, which does not contradict this interpretation because these modes are not subject to the frequency limit.

In the locally Boussinesq but globally anelastic approximation IGWs follow the dispersion relation \citep[e.g.,][]{press1981a},
\begin{equation}
  \label{eq:dispersion-igw}
  \frac{k_\perp}{k} = \frac{\omega}{N}.
\end{equation}
Here, $k_\perp$ is the horizontal wave number, $k_r$ the radial wave number, and $k=\sqrt{k_\perp^2 + k_r^2}$ the magnitude of the total wave vector. We verify this relation for individual values of angular degree~$l$ and frequencies because the resulting velocity field is a combination of many individual waves. The horizontal wave vector can easily be computed for a given $l$ by
\begin{equation}
  k_\perp = \frac{\sqrt{l(l+1)}}{r}.
\end{equation}
The radial wave number is not as straightforward to derive because the wave length changes with radius as the \bvf{} varies. An additional complication is that the wavelength becomes comparable to the stellar radius above $r \sim 0.5\,R_\star$, which makes an accurate determination very hard.

We determine the radial wavelength~$\lambda_r$ for each individual frequency by finding the peaks of $v_r$ along a ray in the radial direction and calculating the difference between them. For this we employ the routine \verb|signal.find_peaks| from the scipy Python package \citep{scipy}, which finds isolated local extrema and is resilient to small numerical noise. As there are only very few wave cycles along the total radius of the star, we use a cubic spline to interpolate the wavelengths at every radial coordinate. Several other methods proved unsuccessful in this particular case: using a radial Fourier transform with a sliding window is inaccurate, as there are only few wave cycles per window; calculating the radial derivative of the phase of the spectrum works well except for regions where the phase is poorly defined when the amplitude is close to 0. This makes this method inapplicable to determining the wavelength in standing modes. The simple method of measuring the distance between peaks and interpolating the found wavelengths is the most robust.

\begin{figure*}
  \centering
  \includegraphics{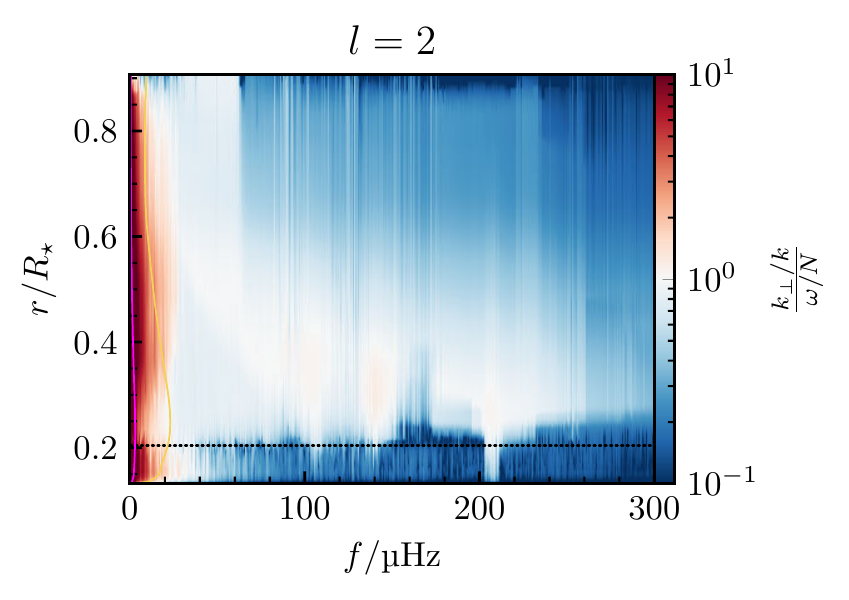}
  \includegraphics{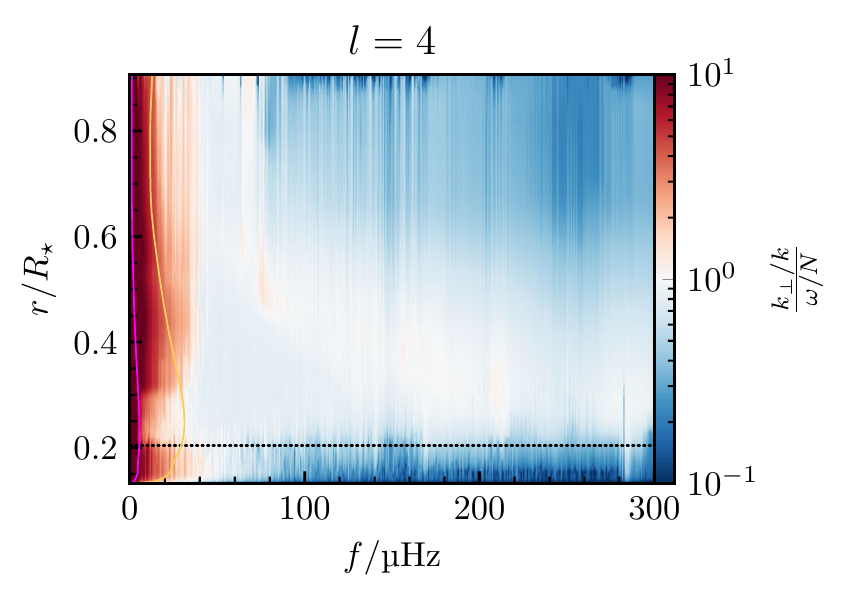}\\
  \includegraphics{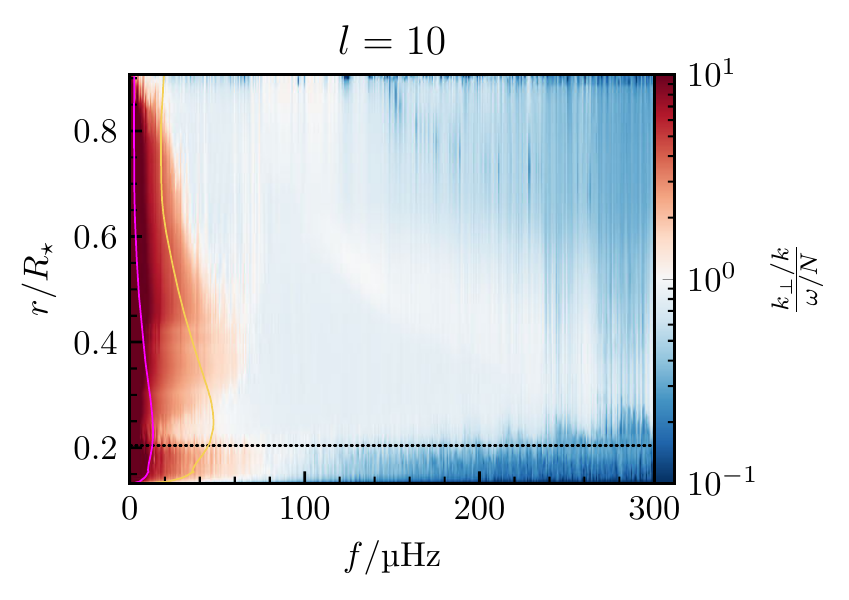}
  \includegraphics{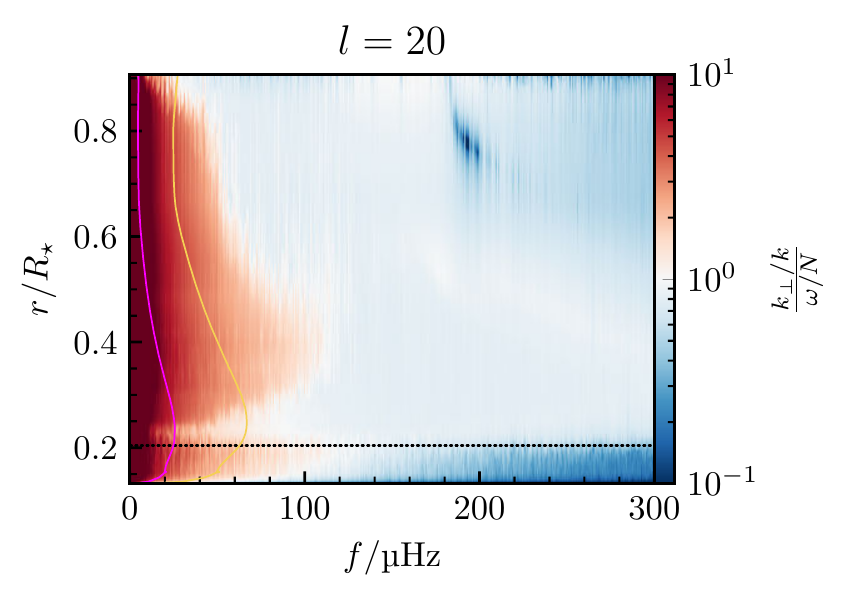}
  \caption{\label{fig:dispersion}Verification of the IGW dispersion relation in Eq.~\eqref{eq:dispersion-igw} for simulation H6LD\@. The color map shows $(k_\perp/k)/(\omega/N)$, which is close to 1 if the relation is fulfilled (white regions). Red regions indicate a too large value of $k_\perp$, blue regions a too low value of $k_\perp$. The magenta lines indicate the region left of which waves are resolved with less than 10 grid points per wave length in radial direction. The yellow line is the estimated upper limit of frequencies dominated by diffusion according to Eq.~\eqref{eq:kappa-flim}. The horizontal dotted line is the boundary of the overshooting region determined in Sect.~\ref{sec:conv-spectra} at $0.695\,H_P$ beyond the original convection zone.}
\end{figure*}

Using this method we calculate the radial wave number $k_r=2\pi/\lambda_r$ for every frequency at every radius to check how closely Eq.~\eqref{eq:dispersion-igw} is fulfilled. The two panels in Fig.~\ref{fig:dispersion} show this for $l=2,4,10,20$. White regions indicate a match of the dispersion relation, red regions have a too large $k_\perp$, blue regions have a too small $k_\perp$. All components show no match in the very low frequency range ($\lesssim \SI{10}{\micro Hz}$), which is expected to be totally dominated by diffusion at all radii. Just above the overshooting region at $0.2\,R_\star$ we find excellent agreement at higher frequencies, which we interpret as waves being emitted from the convection zone over a large range of frequencies. Due to the increased thermal diffusivity needed for numerical reasons, low-frequency waves cannot propagate far into the RZ\@. This is evidenced by the increasing size of the non-IGW (red) region at the low-frequency end.

The higher $l$~values show a remarkable phenomenon. At $r\gtrsim 0.4 \, R_\star$ in the low-frequency region which should be completely dominated by damping ($f\lesssim\SI{70}{\micro Hz}$ for $l=10$) a signal appears which matches the dispersion relation. This cannot be explained by waves originating from the convective boundary because there are no waves of these frequencies present at lower radii. A plausible explanation is that these are generated by nonlinear interaction of low~$l$ waves in the middle of the RZ\@. These secondary waves reach frequencies from \SIrange{10}{100}{\micro Hz}.

The ability of the discretization to resolve IGWs is checked in this context as well. For given values of $k_\perp$ and $N$ we can calculate a frequency below which the radial IGW wavelength would not be resolved by at least 10 grid points in the radial direction. This frequency forms the lower limit for resolving IGWs at a given radius in this simulation. The limit is indicated as a magenta line in Fig.~\ref{fig:dispersion}. The lower limit is highest close to the convective boundary, where $N$ is lowest.

Another limit on wave resolution is imposed by diffusion. As a rough estimate for the minimum wave length of waves not dissipated by diffusion and viscosity we use,
\begin{equation}
  \label{eq:kappa-flim}
  \max(\overline{\kappa}, \overline{\nu}) \sim \frac{\lambda^2}{\tau_\mathrm{d}} = \frac{(2\pi)^5 r^2 f^4}{N^3 l (l+1)},
\end{equation}
with the IGW wavelength~$\lambda=2\pi/k$ and the diffusion time $\tau_\mathrm{d} = \lambda_\perp/v_\mathrm{g}$. This uses the magnitude of the group velocity,
\begin{equation}
  v_g = \frac{\partial\omega}{\partial k} = \frac{r \omega^2}{N\sqrt{l(l+1)}}.
\end{equation}
Solving this equation for $f$ leads to the yellow colored line in Fig.~\ref{fig:dispersion}. As expected, we can see that motions below this frequency largely do not fulfill the dispersion relation. This proves that the radial resolution in our simulations is sufficient to resolve waves with frequencies above \SI{30}{\micro Hz} in the most energy bearing wave numbers ($l \lesssim 5$).

\begin{figure}
  \centering
  \includegraphics{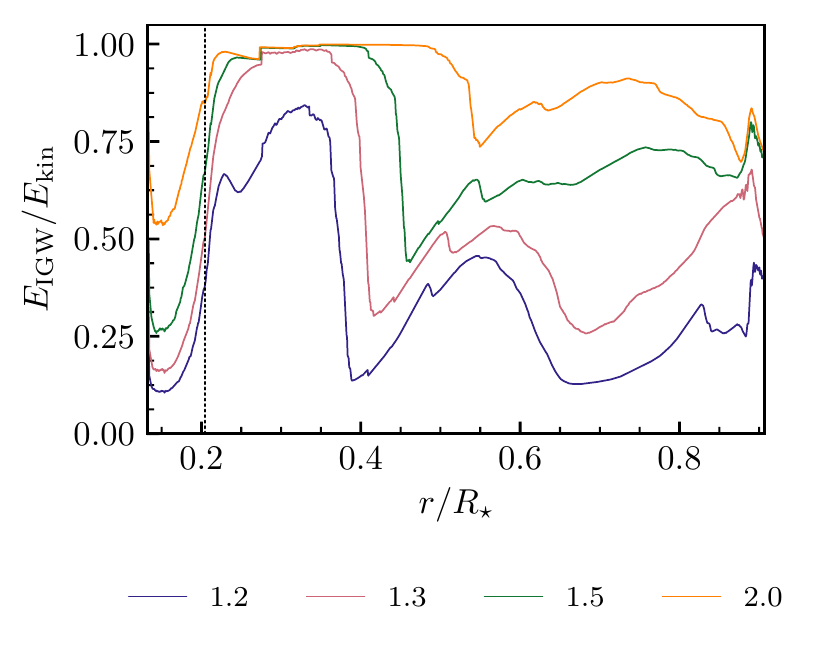}
  \caption{\label{fig:dispersion-filter}Fraction of energy in IGW motions in simulation H6LD after applying the filtering process from Eq.~\eqref{eq:dispersion-filter} at different radii for angular degrees~$l \le 10$. The energy computed from the filtered velocities~$E_\mathrm{IGW}$ is compared to the original kinetic energy~$E_\mathrm{kin}$ for these values of $l$. We do not include frequencies below the limit definitely dominated by diffusion (Eq.~\eqref{eq:kappa-flim}). The line colors represent different error margins~$C_\mathrm{cut}$ around the expected dispersion relation. The vertical dotted line is the boundary of the overshooting region determined in Sect.~\ref{sec:conv-spectra} at $0.695\,H_P$ beyond the original convection zone.}
\end{figure}

While this analysis gives a good overview of the frequencies, wave numbers, and radii where the dispersion relation is fulfilled well, it is also important to see what fraction of the kinetic energy is actually contained in the wave motions. To compute this we filter the Fourier transform of the velocity components $\hat{v}_r$, $\hat{v}_\theta$, and $\hat{v}_\phi$ to only include values at frequencies and radii, where $k_\perp/k$ is close to the IGW dispersion relation $\omega/N$. The filtered velocities are,
\begin{equation}
  \label{eq:dispersion-filter}
  \hat{v}_{r, \theta, \phi}^\mathrm{filt} =
  \begin{cases}
  \hat{v}_{r, \theta, \phi} &\text{if $\frac{\omega}{N} \frac{1}{C_\mathrm{cut}} < \frac{k_\perp}{k} < \frac{\omega}{N} {C_\mathrm{cut}}$,}\\
  0 &\text{otherwise.}
  \end{cases}
\end{equation}
The kinetic energy computed from these velocities is then identified as the energy in IGW motions~$E_\mathrm{IGW}$. Figure~\ref{fig:dispersion-filter} shows the ratio of this energy to the unfiltered kinetic energy integrated over  angular degrees $l \le 10$. We do not include frequencies below the limit definitely dominated by diffusion (Eq.~\eqref{eq:kappa-flim}) in this analysis. As expected, we see almost no energy is in IGWs from the center up to the top of the overshooting region at $r=0.2\,R_\star$, which matches the previously determined position from Sect.~\ref{sec:conv-spectra} as indicated by the vertical dotted line. In the RZ it rises to 90\% when applying the error margin $C_\mathrm{cut}=1.3$. The fraction of kinetic energy in IGWs drops beyond $r=0.4\,R_\star$. The main cause of this is the uncertainty in determining $\lambda_r$ at large radii, where $\lambda_r$ approaches $R_\star$ and our method of measuring the distance between peaks breaks down. Another reason is the growth of the low-frequency, red regions in Fig.~\ref{fig:dispersion}, which arises because of the limited range of IGWs due to high numerical diffusion. At even higher radii, $r \gtrsim 0.7\,R_\star$, we notice an increase in the IGW energy fraction, which is likely due to the increased fraction of secondary waves, as discussed earlier.

This analysis makes us confident that the motions in the RZ are indeed of IGW nature to a significant fraction, with the exception of very low frequency motions dominated by numerical diffusion. We see evidence for secondary generation of waves within the RZ\@.

\subsection{IGW surface signature}
\label{sec:igw-surface}
\begin{figure}
  \includegraphics{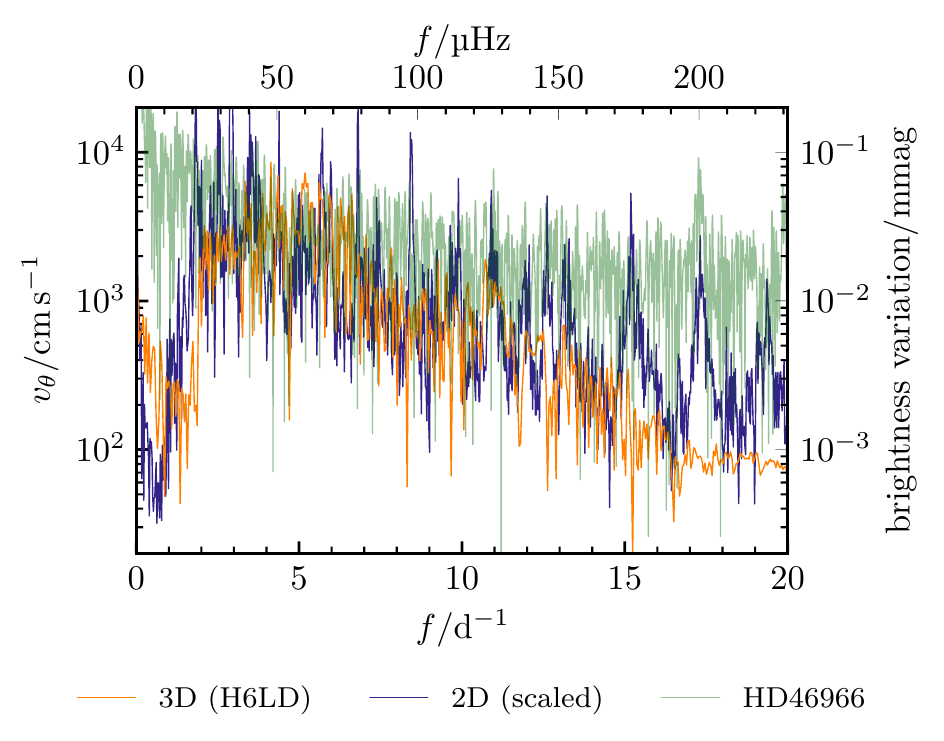}
  \caption{\label{fig:vr-spec-H6LD-surface}Frequency spectrum of tangential velocity at the equator of model~H6E close to the outer boundary of the simulation domain ($r=0.89R_\star$). Just the velocity in $\theta$~direction (i.e., aligned in southern direction at the equator) is used, as the $\phi$~velocity is subject to boundary artifacts. The 2D spectrum from \cite{rogers2013a} was scaled to match the 3D spectrum. It was started with an initially uniform rotation of \SI{1.1}{d^{-1}} using a different, but similar $3\,\msol$ reference state.
    The spectrum of brightness variations of HD46966 is from CoRoT observations \citep{blomme2011a,aerts2015a}. The empirical conversion factor between velocity and brightness variations is \SI{1}{mmag.(km.s^{-1})^{-1}} \citep{decat2002a,aerts2015a}.
  }
\end{figure}

The observed brightness variations in O~stars have been suggested as signatures of convectively excited IGWs \citep{aerts2015a}. Their spectrum is likely linked to that of tangential velocity close to the surface of the star \citep{decat2002a,tkachenko2014a}. Figure~\ref{fig:vr-spec-H6LD-surface} shows a spectrum of latitudinal velocities from model H6LD (\emph{orange line}). These are less affected by numerical influence from the boundary condition than the azimuthal velocities. This is compared to a spectrum obtained from 2D simulations of a $3\,\msol$ star from \citet{rogers2013a} (\emph{blue line}). We see the same low-frequency power excess in range from \SIrange{2}{6}{d^{-1}} and a similar drop in amplitude below \SI{2}{d^{-1}} as in the 3D simulations. We also plot photometric observations \citep{blomme2011a} with an amplitude ratio of \SI{1}{mmag.(km.s^{-1})^{-1}} \citep{decat2002a,aerts2015a} for comparison.

\begin{figure}
  \includegraphics{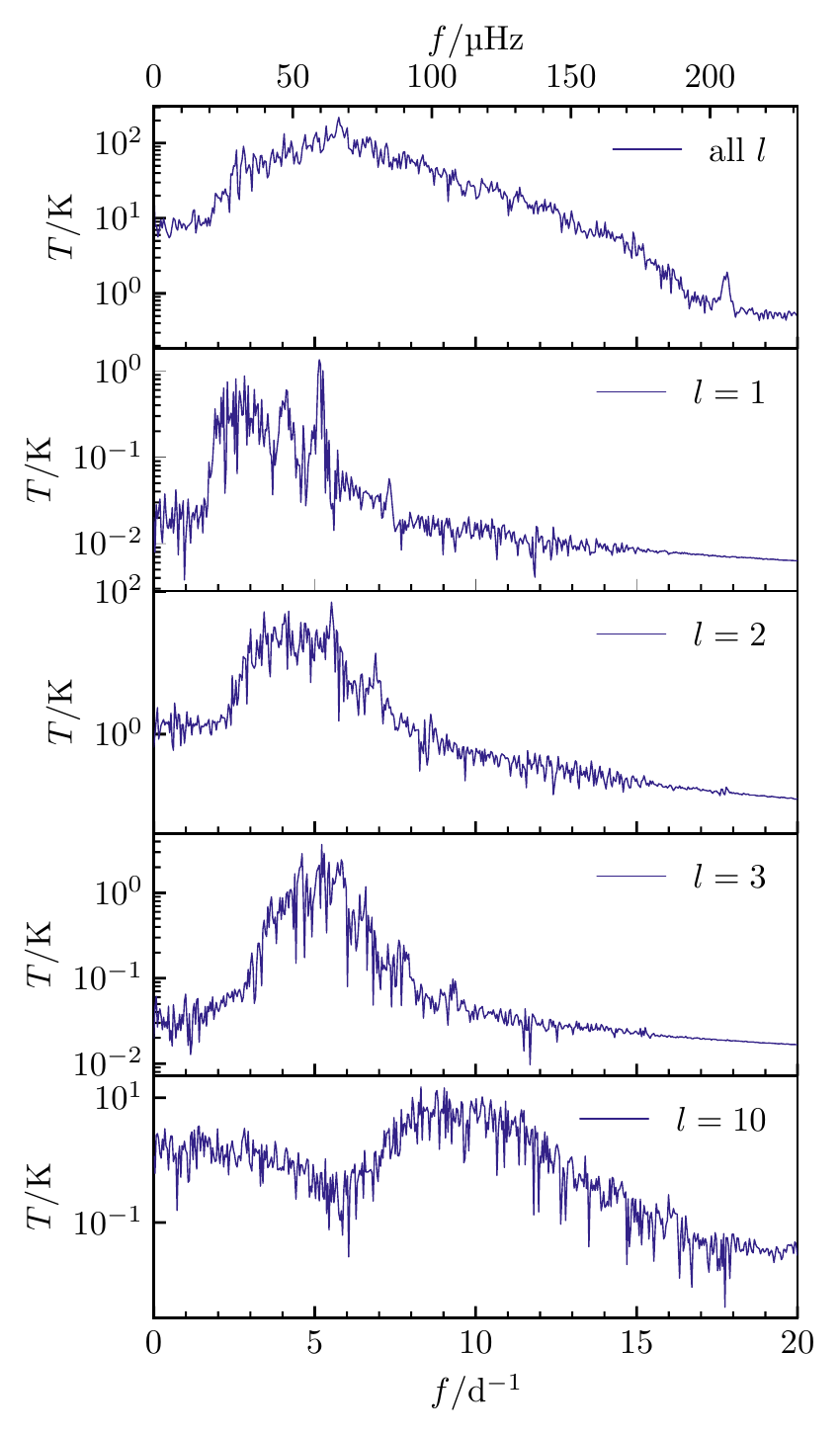}
  \caption{\label{fig:spec-H6LD-surface-T}Frequency spectrum of temperature fluctuations~$T$ from the reference state at the equator of model~H6LD close to the outer boundary of the simulation domain ($r=0.89\;R_\star$). The different panels show different values of angular degree ($l$) as indicated.}
\end{figure}

The same low-frequency power excess shows in the spectra of temperature fluctuations in Fig.~\ref{fig:spec-H6LD-surface-T}. It is expected to be the dominant cause of photometric variability in observations. In the spectrum integrated over all $l$ values (top panel) this excess makes it hard to distinguish individual excited mode frequencies, expect for one mode at \SI{210}{\micro Hz} (\SI{18.1}{d^{-1}}), which is part of the $l=2$ component. In the spectra for individual $l$ values, several modes can be identified, corresponding to those in Fig.~\ref{fig:vr-spec-pcolor-l-2-4}. This decomposition also makes it clear that the low-frequency power excess is a combination of the power excesses in different $l$ components, each contributing to a small frequency range. The lack of signal at low-frequencies in numerical simulations is due to the high numerical diffusivity and not expected to be physical. The amplitude of the waves is expected to increase as they propagate from $r=0.89\;R_\star$, where the spectrum was computed, to the surface. According to pseudomomentum conservation it should increase by a factor of 380.

The low-frequency power excess is also found in observations of stars with a convective core \citep{bowman2019a}. In the simulations it is caused by the high density of high radial order, low-$l$ g~modes and because most energy in the CZ is at low $l$~values (see Fig.~\ref{fig:k-spectrum}). The drop in amplitude below \SI{2}{d^{-1}} on the other hand is in disagreement with observed photometry \citep{blomme2011a,aerts2017a,aerts2018a,bowman2019a}. This disagreement is likely caused by the increased thermal diffusivity in the simulations (both in 2D and 3D) which damps low-frequency waves more strongly than in stellar interiors. Another possibility is the lack of differential rotation in our 3D model. \citet{rogers2013a} found that differential rotation between core and envelope introduces a significant low frequency component in the spectrum. At higher frequencies above \SI{10}{d^{-1}} the 2D simulations drop more slowly than the 3D simulations and show many excited modes. This is possibly due to the lack of wave breaking brought about by the high thermal diffusivity and viscosity needed in our present set of 3D simulations.

\section{Conclusions and Outlook} \label{sec:conclusions}
We showed the first 3D simulations of convection in the core of an intermediate-mass star, with a convective core and radiative envelope, that also include a large part of the radiation zone (RZ). The simulations using the anelastic equations (i.e., removing the physics of sound waves) and a spectral discretization using spherical harmonics were run using a realistic reference state from the stellar evolution code MESA\@. For numerical reasons the simulations were run with increased thermal and viscous diffusivity. To compensate for the increased wave damping this produces we increased the luminosity of the star causing higher velocities in the convective core. We do this in the hope that wave velocities at the surface of the star are more realistic.

We see wave patterns in the RZ, which are identified to be standing g and f~modes with frequencies similar to those predicted by the oscillation code GYRE\@. Although there are differences, they are not of the sort predicted in \citet{brown2012a} and are dependent on the $l$ and $m$ values of the spherical harmonics. These differences are likely due to slightly different physics (e.g., equation of state, outer boundary condition) and changes in the temperature profile at the top of the convection zone due to overshooting.

Apart from the standing modes the simulations also show a continuous signal in the RZ between frequencies of approximately \SI{20}{\micro Hz} and \SI{200}{\micro Hz}. An analysis of the dispersion relation (see Fig.~\ref{fig:dispersion}) identifies the physical mechanism as IGWs. The decline of this continuous spectrum with frequency is markedly smaller than theoretically predicted values for excitation purely due to convective eddies (see Fig.~\ref{fig:H6LD-spec-rads}).

An analysis of the kinetic energy distribution over spherical harmonic degree~$l$ shows a spectrum which peaks at a low value of $l$ and then declines with a power law with an exponent in the range from \numrange{-2}{-3} in the inertial range. This is closer to the theoretical value for Bolgiano--Obukhov scaling ($-2.2$) of buoyancy dominated convection than to the Kolmogorov value ($-1.6$) of isotropic turbulence. We do not have enough information to get a conclusive answer on the realized scaling in convective stellar cores. This should be studied further in detailed simulations of just the core. These slopes are measured at the top of the CZ, which is subject to convective overshooting, and therefore do not directly match the IGW spectrum. Yet they show what energy is available for wave excitation at a given frequency. The slope in the inertial range is similar to that observed in other 2D or 3D simulations of core convection \citep{rogers2013a,augustson2016a}.

The broken power law structure of the frequency spectra of kinetic energy above the convection zone is similar to those obtained in the 2D simulations of \citet{rogers2013a}, suggesting the mechanism driving the bulk of this spectrum does not fundamentally change with dimensionality.
It is likely that bulk Reynolds stresses induced by convective eddies contribute more in higher Reynolds number flows, but this would still only affect the low frequencies ($f<f_\text{TO}$) and hence, have little impact on angular momentum transport or mixing within the bulk of the RZ \citep{shiode2013a,kumar1999a,lecoanet2013a}.

Excitation by plume penetration is obviously involved as can be seen in the temperature and velocity fields (see Fig.~\ref{fig:H6R10-slice}). It can explain the excitation of higher frequency waves and the extracted distribution of plume frequencies fits a large part of the simulation spectrum. One may argue that the plume penetration depths, and hence, frequencies generated are too large. However, at least in the theory by \citet{zahn1991a}, this penetration depth scales like the velocities cubed divided by the total flux, a number which is the same in the simulations as it is in the star. The production of high-frequency waves is extremely important for explaining the photometrically observed brightness variations at high frequencies (see Fig.~\ref{fig:vr-spec-H6LD-surface}). They are likely underestimated in our simulations due to high dissipation preventing wave breaking.

Stochastic brightness variations caused by velocity and temperature fluctuations at the stellar surface have been inferred to be caused by IGWs in massive stars \citep{aerts2015a,aerts2017b,aerts2018a,bowman2019a}. We extracted frequency spectra of these quantities from the simulation close to the stellar surface. General features are a low-frequency power excess and the presence of standing modes at low $l$ harmonics. This is in agreement with the findings of 2D simulations \citep{rogers2013a}, which match observations in the power bearing range, but lack both amplitudes at very low frequencies (due to excessive radiative damping) and high frequencies (possibly due to lack of wave breaking from overdamped waves).

The simulations presented in this article show the feasibility of hydrodynamic modeling of convectively excited IGWs and their propagation through a large part of the radiative zone using a consistent numerical treatment. In future work employing more computational resources the limitations forcing us to use unphysically high diffusivities and luminosities can hopefully be overcome to achieve more realistic wave amplitudes throughout the interior and at the surface, and hence more realistic angular momentum transport. More realistic physical parameters in the simulations combined with coverage of a wider range of stellar models will also allow us to make quantitative predictions of the expected signature of IGWs in asteroseismological observations.

The general similarity of our results with those of previous 2D simulations encourage us to consider those results with less reservations due to their dimensionality and to use 2D simulations as a tool in the future to quickly cover a wider parameter range in models than is possible with 3D simulations.

\acknowledgments
Support for this research was provided by STFC grant ST/L005549/1 and NASA grant NNX17AB92G.
MGP and DMB received funding from the European Research Council (ERC) under the European Union's Horizon 2020 research and innovation programme (grant agreement No670519: MAMSIE).
VP acknowledges support from the European Research Council through ERC grant SPIRE 647383.
Resources supporting this work were provided by the NASA High-End Computing (HEC) Program through the NASA Advanced Supercomputing (NAS) Division at Ames Research Center.
This research made use of the Rocket High Performance Computing service at Newcastle University. The authors thank C.~Pin{\c c}on and M.~Rieutord for helpful comments.
\vspace{5mm}

\software{matplotlib \citep{hunter2007a}, scipy \citep{scipy}, MESA \citep{paxton2011a,paxton2013a,paxton2015a,paxton2018a}, GYRE \citep{townsend2013a}}

\bibliography{igw3d}

\begin{thebibliography}{}
\expandafter\ifx\csname natexlab\endcsname\relax\def\natexlab#1{#1}\fi
\providecommand{\url}[1]{\href{#1}{#1}}

\bibitem[{{Aerts} {et~al.}(2010){Aerts}, {Christensen-Dalsgaard}, \&
  {Kurtz}}]{aerts2010a}
{Aerts}, C., {Christensen-Dalsgaard}, J., \& {Kurtz}, D.~W. 2010,
  {Asteroseismology}

\bibitem[{{Aerts} \& {Rogers}(2015)}]{aerts2015a}
{Aerts}, C., \& {Rogers}, T.~M. 2015, \apjl, 806, L33

\bibitem[{{Aerts} {et~al.}(2017{\natexlab{a}}){Aerts}, {Van Reeth}, \&
  {Tkachenko}}]{aerts2017b}
{Aerts}, C., {Van Reeth}, T., \& {Tkachenko}, A. 2017{\natexlab{a}}, \apjl,
  847, L7

\bibitem[{{Aerts} {et~al.}(2017{\natexlab{b}}){Aerts},
  {S{\'{\i}}mon-D{\'{\i}}az}, {Bloemen}, {Debosscher}, {P{\'a}pics}, {Bryson},
  {Still}, {Moravveji}, {Williamson}, {Grundahl}, {Fredslund Andersen},
  {Antoci}, {Pall{\'e}}, {Christensen-Dalsgaard}, \& {Rogers}}]{aerts2017a}
{Aerts}, C., {S{\'{\i}}mon-D{\'{\i}}az}, S., {Bloemen}, S., {et~al.}
  2017{\natexlab{b}}, \aap, 602, A32

\bibitem[{{Aerts} {et~al.}(2018){Aerts}, {Bowman}, {S{\'{\i}}mon-D{\'{\i}}az},
  {Buysschaert}, {Johnston}, {Moravveji}, {Beck}, {De Cat}, {Triana},
  {Aigrain}, {Castro}, {Huber}, \& {White}}]{aerts2018a}
{Aerts}, C., {Bowman}, D.~M., {S{\'{\i}}mon-D{\'{\i}}az}, S., {et~al.} 2018,
  \mnras, 476, 1234

\bibitem[{{Alvan} {et~al.}(2014){Alvan}, {Brun}, \& {Mathis}}]{alvan2014a}
{Alvan}, L., {Brun}, A.~S., \& {Mathis}, S. 2014, \aap, 565, A42

\bibitem[{{Augustson} {et~al.}(2016){Augustson}, {Brun}, \&
  {Toomre}}]{augustson2016a}
{Augustson}, K.~C., {Brun}, A.~S., \& {Toomre}, J. 2016, \apj, 829, 92

\bibitem[{{Baldwin} {et~al.}(2001){Baldwin}, {Gray}, {Dunkerton}, {Hamilton},
  {Haynes}, {Randel}, {Holton}, {Alexander}, {Hirota}, {Horinouchi}, {Jones},
  {Kinnersley}, {Marquardt}, {Sato}, \& {Takahashi}}]{baldwin2001a}
{Baldwin}, M.~P., {Gray}, L.~J., {Dunkerton}, T.~J., {et~al.} 2001, Reviews of
  Geophysics, 39, 179

\bibitem[{{Beck} {et~al.}(2012){Beck}, {Montalban}, {Kallinger}, {De Ridder},
  {Aerts}, {Garc{\'{\i}}a}, {Hekker}, {Dupret}, {Mosser}, {Eggenberger},
  {Stello}, {Elsworth}, {Frandsen}, {Carrier}, {Hillen}, {Gruberbauer},
  {Christensen-Dalsgaard}, {Miglio}, {Valentini}, {Bedding}, {Kjeldsen},
  {Girouard}, {Hall}, \& {Ibrahim}}]{beck2012a}
{Beck}, P.~G., {Montalban}, J., {Kallinger}, T., {et~al.} 2012, \nat, 481, 55

\bibitem[{{Blomme} {et~al.}(2011){Blomme}, {Mahy}, {Catala}, {Cuypers},
  {Gosset}, {Godart}, {Montalban}, {Ventura}, {Rauw}, {Morel}, {Degroote},
  {Aerts}, {Noels}, {Michel}, {Baudin}, {Baglin}, {Auvergne}, \&
  {Samadi}}]{blomme2011a}
{Blomme}, R., {Mahy}, L., {Catala}, C., {et~al.} 2011, \aap, 533, A4

\bibitem[{{Bolgiano}(1959)}]{bolgiano1959a}
{Bolgiano}, Jr., R. 1959, \jgr, 64, 2226

\bibitem[{{Bowman} {et~al.}(2019){Bowman}, {Aerts}, {Johnston}, {Pedersen},
  {Rogers}, {Edelmann}, {Sim{\'o}n-D{\'{\i}}az}, {Van Reeth}, {Buysschaert},
  {Tkachenko}, \& {Triana}}]{bowman2019a}
{Bowman}, D.~M., {Aerts}, C., {Johnston}, C., {et~al.} 2019, \aap, 621, A135

\bibitem[{{Braginsky} \& {Roberts}(1995)}]{braginsky1995a}
{Braginsky}, S.~I., \& {Roberts}, P.~H. 1995, Geophysical and Astrophysical
  Fluid Dynamics, 79, 1

\bibitem[{{Brown} {et~al.}(2012){Brown}, {Vasil}, \& {Zweibel}}]{brown2012a}
{Brown}, B.~P., {Vasil}, G.~M., \& {Zweibel}, E.~G. 2012, \apj, 756, 109

\bibitem[{{Browning} {et~al.}(2004){Browning}, {Brun}, \&
  {Toomre}}]{browning2004a}
{Browning}, M.~K., {Brun}, A.~S., \& {Toomre}, J. 2004, \apj, 601, 512

\bibitem[{{Brun} {et~al.}(2005){Brun}, {Browning}, \& {Toomre}}]{brun2005a}
{Brun}, A.~S., {Browning}, M.~K., \& {Toomre}, J. 2005, \apj, 629, 461

\bibitem[{{B{\"u}hler}(2009)}]{buehler2009a}
{B{\"u}hler}, O. 2009, {Waves and Mean Flows} (Cambridge University Press)

\bibitem[{Clune {et~al.}(1999)Clune, Elliott, Miesch, Toomre, \&
  Glatzmaier}]{clune1999a}
Clune, T.~C., Elliott, J.~R., Miesch, M.~S., Toomre, J., \& Glatzmaier, G.~A.
  1999, Parallel Computing, 25, 361

\bibitem[{{Cristini} {et~al.}(2017){Cristini}, {Meakin}, {Hirschi}, {Arnett},
  {Georgy}, {Viallet}, \& {Walkington}}]{cristini2017a}
{Cristini}, A., {Meakin}, C., {Hirschi}, R., {et~al.} 2017, \mnras, 471, 279

\bibitem[{{De Cat} \& {Aerts}(2002)}]{decat2002a}
{De Cat}, P., \& {Aerts}, C. 2002, \aap, 393, 965

\bibitem[{{Denissenkov} {et~al.}(2008){Denissenkov}, {Pinsonneault}, \&
  {MacGregor}}]{denissenkov2008a}
{Denissenkov}, P.~A., {Pinsonneault}, M., \& {MacGregor}, K.~B. 2008, \apj,
  684, 757

\bibitem[{{Freytag} {et~al.}(1996){Freytag}, {Ludwig}, \&
  {Steffen}}]{freytag1996a}
{Freytag}, B., {Ludwig}, H.-G., \& {Steffen}, M. 1996, \aap, 313, 497

\bibitem[{{Frisch}(1995)}]{frisch1995a}
{Frisch}, U. 1995, Turbulence (Cambridge: Cambridge Universtiy Press)

\bibitem[{{Fuller} {et~al.}(2014){Fuller}, {Lecoanet}, {Cantiello}, \&
  {Brown}}]{fuller2014a}
{Fuller}, J., {Lecoanet}, D., {Cantiello}, M., \& {Brown}, B. 2014, \apj, 796,
  17

\bibitem[{{Garcia Lopez} \& {Spruit}(1991)}]{garcia-lopez1991a}
{Garcia Lopez}, R.~J., \& {Spruit}, H.~C. 1991, \apj, 377, 268

\bibitem[{{Glatzmaier}(1984)}]{glatzmaier1984a}
{Glatzmaier}, G.~A. 1984, Journal of Computational Physics, 55, 461

\bibitem[{{Glatzmaier}(2013)}]{glatzmaier2013a}
---. 2013, {Introduction to Modelling Convection in Planets and Stars}

\bibitem[{{Goldreich} \& {Kumar}(1990)}]{goldreich1990a}
{Goldreich}, P., \& {Kumar}, P. 1990, \apj, 363, 694

\bibitem[{Hunter(2007)}]{hunter2007a}
Hunter, J.~D. 2007, Computing In Science \& Engineering, 9, 90

\bibitem[{Jones {et~al.}(2001--)Jones, Oliphant, Peterson, {et~al.}}]{scipy}
Jones, E., Oliphant, T., Peterson, P., {et~al.} 2001--, {SciPy}: Open source
  scientific tools for {Python}, , .
\newblock \url{http://www.scipy.org/}

\bibitem[{{Jones} {et~al.}(2017){Jones}, {Andrassy}, {Sandalski}, {Davis},
  {Woodward}, \& {Herwig}}]{jones2017a}
{Jones}, S., {Andrassy}, R., {Sandalski}, S., {et~al.} 2017, \mnras, 465, 2991

\bibitem[{{Kippenhahn} {et~al.}(2012){Kippenhahn}, {Weigert}, \&
  {Weiss}}]{kippenhahn2012a}
{Kippenhahn}, R., {Weigert}, A., \& {Weiss}, A. 2012, {Stellar Structure and
  Evolution} (Berlin Heidelberg: Springer-Verlag),
  doi:10.1007/978-3-642-30304-3

\bibitem[{{Kolmogorov}(1941)}]{kolmogorov1941a}
{Kolmogorov}, A.~N. 1941, Dokl.~Akad.~Nauk SSSR, 30, 299, in Russian

\bibitem[{{Kumar} {et~al.}(1999){Kumar}, {Talon}, \& {Zahn}}]{kumar1999a}
{Kumar}, P., {Talon}, S., \& {Zahn}, J.-P. 1999, \apj, 520, 859

\bibitem[{{Lecoanet} \& {Quataert}(2013)}]{lecoanet2013a}
{Lecoanet}, D., \& {Quataert}, E. 2013, \mnras, 430, 2363

\bibitem[{{Lighthill}(1952)}]{lighthill1952a}
{Lighthill}, M.~J. 1952, Proceedings of the Royal Society of London Series A,
  211, 564

\bibitem[{{Meakin} \& {Arnett}(2007)}]{meakin2007a}
{Meakin}, C.~A., \& {Arnett}, D. 2007, \apj, 667, 448

\bibitem[{{Montalb{\'a}n}(1994)}]{montalban1994a}
{Montalb{\'a}n}, J. 1994, \aap, 281, 421

\bibitem[{{Montalb{\'a}n} \& {Schatzman}(2000)}]{montalban2000a}
{Montalb{\'a}n}, J., \& {Schatzman}, E. 2000, \aap, 354, 943

\bibitem[{{Munk} \& {Wunsch}(1998)}]{munk1998a}
{Munk}, W., \& {Wunsch}, C. 1998, Deep Sea Research Part I: Oceanographic
  Research, 45, 1977

\bibitem[{Obukhov(1959)}]{obukhov1959a}
Obukhov, A. 1959, in Dokl. Akad. Nauk. SSSR, Vol. 125, 1246

\bibitem[{{Paxton} {et~al.}(2011){Paxton}, {Bildsten}, {Dotter}, {Herwig},
  {Lesaffre}, \& {Timmes}}]{paxton2011a}
{Paxton}, B., {Bildsten}, L., {Dotter}, A., {et~al.} 2011, \apjs, 192, 3

\bibitem[{{Paxton} {et~al.}(2013){Paxton}, {Cantiello}, {Arras}, {Bildsten},
  {Brown}, {Dotter}, {Mankovich}, {Montgomery}, {Stello}, {Timmes}, \&
  {Townsend}}]{paxton2013a}
{Paxton}, B., {Cantiello}, M., {Arras}, P., {et~al.} 2013, \apjs, 208, 4

\bibitem[{{Paxton} {et~al.}(2015){Paxton}, {Marchant}, {Schwab}, {Bauer},
  {Bildsten}, {Cantiello}, {Dessart}, {Farmer}, {Hu}, {Langer}, {Townsend},
  {Townsley}, \& {Timmes}}]{paxton2015a}
{Paxton}, B., {Marchant}, P., {Schwab}, J., {et~al.} 2015, \apjs, 220, 15

\bibitem[{{Paxton} {et~al.}(2018){Paxton}, {Schwab}, {Bauer}, {Bildsten},
  {Blinnikov}, {Duffell}, {Farmer}, {Goldberg}, {Marchant}, {Sorokina},
  {Thoul}, {Townsend}, \& {Timmes}}]{paxton2018a}
{Paxton}, B., {Schwab}, J., {Bauer}, E.~B., {et~al.} 2018, \apjs, 234, 34

\bibitem[{{Pin{\c c}on} {et~al.}(2016){Pin{\c c}on}, {Belkacem}, \&
  {Goupil}}]{pincon2016a}
{Pin{\c c}on}, C., {Belkacem}, K., \& {Goupil}, M.~J. 2016, \aap, 588, A122

\bibitem[{{Porter} \& {Woodward}(2000)}]{porter2000a}
{Porter}, D.~H., \& {Woodward}, P.~R. 2000, \apjs, 127, 159

\bibitem[{{Press}(1981)}]{press1981a}
{Press}, W.~H. 1981, \apj, 245, 286

\bibitem[{{Quataert} \& {Shiode}(2012)}]{quataert2012a}
{Quataert}, E., \& {Shiode}, J. 2012, \mnras, 423, L92

\bibitem[{{Ratnasingam} {et~al.}(2019){Ratnasingam}, {Edelmann}, \&
  {Rogers}}]{ratnasingam2019a}
{Ratnasingam}, R.~P., {Edelmann}, P.~V.~F., \& {Rogers}, T.~M. 2019, \mnras,
  482, 5500

\bibitem[{{Rieutord} \& {Zahn}(1995)}]{rieutord1995a}
{Rieutord}, M., \& {Zahn}, J.-P. 1995, \aap, 296, 127

\bibitem[{{Rogers}(2015)}]{rogers2015a}
{Rogers}, T.~M. 2015, \apjl, 815, L30

\bibitem[{{Rogers} \& {Glatzmaier}(2005)}]{rogers2005b}
{Rogers}, T.~M., \& {Glatzmaier}, G.~A. 2005, \mnras, 364, 1135

\bibitem[{{Rogers} {et~al.}(2006){Rogers}, {Glatzmaier}, \&
  {Jones}}]{rogers2006a}
{Rogers}, T.~M., {Glatzmaier}, G.~A., \& {Jones}, C.~A. 2006, \apj, 653, 765

\bibitem[{{Rogers} {et~al.}(2013){Rogers}, {Lin}, {McElwaine}, \&
  {Lau}}]{rogers2013a}
{Rogers}, T.~M., {Lin}, D.~N.~C., {McElwaine}, J.~N., \& {Lau}, H.~H.~B. 2013,
  \apj, 772, 21

\bibitem[{{Schatzman}(1993)}]{schatzman1993a}
{Schatzman}, E. 1993, \aap, 279, 431

\bibitem[{{Shiode} {et~al.}(2013){Shiode}, {Quataert}, {Cantiello}, \&
  {Bildsten}}]{shiode2013a}
{Shiode}, J.~H., {Quataert}, E., {Cantiello}, M., \& {Bildsten}, L. 2013,
  \mnras, 430, 1736

\bibitem[{{Talon} \& {Charbonnel}(2005)}]{talon2005a}
{Talon}, S., \& {Charbonnel}, C. 2005, \aap, 440, 981

\bibitem[{{Tkachenko} {et~al.}(2014){Tkachenko}, {Degroote}, {Aerts},
  {Pavlovski}, {Southworth}, {P{\'a}pics}, {Moravveji}, {Kolbas}, {Tsymbal},
  {Debosscher}, \& {Cl{\'e}mer}}]{tkachenko2014a}
{Tkachenko}, A., {Degroote}, P., {Aerts}, C., {et~al.} 2014, \mnras, 438, 3093

\bibitem[{{Townsend}(1966)}]{townsend1966a}
{Townsend}, A.~A. 1966, Journal of Fluid Mechanics, 24, 307

\bibitem[{{Townsend} \& {Teitler}(2013)}]{townsend2013a}
{Townsend}, R.~H.~D., \& {Teitler}, S.~A. 2013, \mnras, 435, 3406

\bibitem[{{Triana} {et~al.}(2015){Triana}, {Moravveji}, {P{\'a}pics}, {Aerts},
  {Kawaler}, \& {Christensen-Dalsgaard}}]{triana2015a}
{Triana}, S.~A., {Moravveji}, E., {P{\'a}pics}, P.~I., {et~al.} 2015, \apj,
  810, 16

\bibitem[{{Viallet} {et~al.}(2013){Viallet}, {Meakin}, {Arnett}, \&
  {Moc{\'a}k}}]{viallet2013a}
{Viallet}, M., {Meakin}, C., {Arnett}, D., \& {Moc{\'a}k}, M. 2013, \apj, 769,
  1

\bibitem[{{Zahn}(1991)}]{zahn1991a}
{Zahn}, J.-P. 1991, \aap, 252, 179

\end{thebibliography}

\end{document}